\documentclass[11pt,letterpaper]{article}

\def\includetimestamp{0}
\def\smallheadings{0}

\usepackage{times,mysetup,my-acm,xspace}

\FULLPAGE

\usepackage[letterpaper,dvips,margin=1in]{geometry}
\usepackage{url}

\ifnum\smallheadings=1
\makeatletter
\def\section{\@startsection{section}{1}{0mm}{-2.0ex plus -0.5ex minus -.1ex}{0.7ex plus .5ex minus .2ex}{\bf\large}}
\def\subsection{\@startsection{subsection}{2}{\z@}{-1.0ex plus -0.5ex minus -.2ex}{0.5ex plus .2ex minus .2ex}{\bf\large}}
\def\subsubsection{\@startsection{subsubsection}{3}{\z@}{-1.0ex plus -0.5ex minus -.2ex}{0.5ex plus .2ex minus .2ex}{\normalsize\bf}}
\def\paragraph{\@startsection{paragraph}{4}{\z@}{1.0ex plus 0.5ex minus .2ex}{-1em}{\normalsize\bf}}
\def\subparagraph{\@startsection {subparagraph}{5}{\parindent}{1.0ex plus 0.5ex minus .2ex}{-1em}{\normalsize\bf}}
\makeatother
\fi

\usepackage{wrapfig}

\ifnum\includetimestamp=1
	\usepackage{datetime}
	 \newdateformat{dashdate}{\THEYEAR-\twodigit{\THEMONTH}-\twodigit{\THEDAY}}
	\newtimeformat{colontime}{\twodigit{\THEHOUR}:\twodigit{\THEMINUTE}}
	 \newcommand{\timestamp}{\dashdate\today-\shortdayofweekname{\THEDAY}{\THEMONTH}{\THEYEAR}-\colontime}
\fi
\usepackage{fancyhdr}

\providecommand{\Appendix}{}
\renewcommand{\Appendix}[2][?]{%
	\refstepcounter{section}%
	\vspace{\parskip}%
	{\flushright
	\ifnum\smallheadings=1 \large \else \Large \fi
	\bfseries\appendixname\ \thesection: #1}%
	\vspace{0.5\baselineskip}%
}

\renewcommand{\appendix}{%
	\renewcommand{\section}{\secdef\Appendix\Appendix}%
	\renewcommand{\thesection}{\Alph{section}}%
	\setcounter{section}{0}%
}
\newcommand{\yoginote}[1]{\sidenote{Yogi}{#1}}

\usepackage{enumerate, paralist, comment}

\usepackage{graphicx} % needed to include figures to compile with pdflatex

\allowdisplaybreaks

\newcommand{\realization}{click realization\xspace}
\newcommand{\realizations}{click realizations\xspace}

\title{{\bf Characterizing Truthful Multi-Armed Bandit Mechanisms}%
\footnote{This is a full version of a conference paper published in \emph{10th ACM Conf. on Electronic Commerce (EC)}, 2009.
Apart from the revised presentation, this version is updated to reflect the follow-up work~\cite{Transform-ec10,SingleCall-ec12,Gatti-ec12,Parkes-netecon12} and the current snapshot of open questions. }}
\author{
{\Large Moshe Babaioff}\\ Microsoft Research Silicon Valley\\ Mountain View, CA 94043, USA\\ moshe@microsoft.com
\and {\Large Yogeshwer Sharma%
\thanks{This research was done while Y. Sharma was a student at Cornell University and an intern at Microsoft Research Silicon Valley.}}\\
Facebook\\ Palo Alto, CA 94301, USA
    \\ yogeshwersharma@gmail.com
\and {\Large Aleksandrs Slivkins}\\ Microsoft Research  Silicon Valley\\ Mountain View, CA 94043, USA\\ slivkins@microsoft.com}
\date{December 2008\\ This version: May 2013}

\begin{document}

\ifnum\includetimestamp=1
	\lhead{}
	\chead{}
	\rhead{}
	\lfoot{}
	\cfoot{\thepage}
	\rfoot{\timestamp}
	\renewcommand{\headrulewidth}{0.2pt}
	\renewcommand{\footrulewidth}{0.2pt}
	\pagestyle{fancy}
\else
	\pagestyle{plain}
\fi

\numberwithin{equation}{section}

\maketitle

\begin{abstract}

We consider a multi-round auction setting motivated by pay-per-click auctions for Internet advertising. In each round the auctioneer selects an advertiser and shows her ad, which is then either clicked or not. An advertiser derives value from clicks; the value of a click is her private information. Initially, neither the auctioneer nor the advertisers have any information about the likelihood of clicks on the advertisements. The auctioneer's goal is to design a (dominant strategies) truthful mechanism that (approximately) maximizes the social welfare.

If the advertisers bid their true private values, our problem is equivalent to the \emph{multi-armed bandit problem}, and thus can be viewed as a strategic version of the latter. In particular, for both problems the quality of an algorithm can be characterized by \emph{regret}, the difference in social welfare between the algorithm and the benchmark which always selects the same ``best'' advertisement. We investigate how the design of multi-armed bandit algorithms is affected by the restriction that the resulting mechanism must be truthful. We find that deterministic truthful mechanisms have certain strong structural properties -- essentially, they must separate exploration from exploitation -- \emph{and} they incur much higher regret than the optimal multi-armed bandit algorithms. Moreover, we provide a truthful mechanism which (essentially) matches our lower bound on regret.
\end{abstract}

{\bf ACM Categories and subject descriptors:}

\OMIT{ %%% replaced the categories as per EC-camera-ready
F.2.2 [nonnumerical algorithms and problems]: computations on discrete structures;~~
J.4 [social and behavioral sciences]: Economics;~~
F.1.2 [modes of computation]: online computation.
}

{\bf General Terms:} theory, algorithms, economics.


\newcommand{\fakeItem}[1][$\bullet$]{\vspace{2mm}\noindent{\bf #1}~~}

\newcommand{\calG}{\mathcal{G}}
\newcommand{\calP}{\mathcal{P}}
\newcommand{\E}{\expectation}
\renewcommand{\P}{\probability}
\newcommand{\explain}[1]{\quad\text{#1}}
\newcommand{\calE}{\mathcal{E}}
\newcommand{\calM}{\mathcal{M}}
\newcommand{\UCB}{\ensuremath{\mathtt{UCB1}}}
\newcommand{\EXP}{\ensuremath{\mathtt{EXP3}}}
\newcommand{\mubar}{\bar\mu}

\newcommand{\Vmax}{\ensuremath{v_{\max}}}
\newcommand{\problem}{MAB mechanism design problem}
\newcommand{\vmax}{\Vmax}
\newcommand{\ctr}{\mu}
\newcommand{\avgctr}{\bar{\mu}}
\newcommand{\ibar}{\bar{i}}
\newcommand{\given}{\,|\,}

\renewcommand{\d}{\,d}
\newcommand{\thatis}{i.e.\ }
\newcommand{\threshold}{\Theta}
\newcommand{\defequal}{\stackrel{\mathrm{def}}{=}}
\newcommand{\Reals}{\mathbb{R}}
\newcommand{\calF}{\mathcal{F}}
\newcommand{\kld}{\mathtt{KLD}}
\newcommand{\expectation}{\mathbb{E}}
\newcommand{\probability}{\mathbb{P}}
\newcommand{\event}{\mathrm{\mathcal{E}}}
\newcommand{\Ttwothird}{T^{2/3}}
\newcommand{\ttwothird}{\Ttwothird}
\newcommand{\half}{\frac{1}{2}}
\newcommand{\bigo}{\mathcal{O}}
\newcommand{\bigomega}{\Omega}
\newcommand{\clean}{\text{clean}}
\newcommand{\dirty}{\text{dirty}}

\newcommand{\adv}{\text{ADV}}
\newcommand{\alg}{\text{ALG}}
\newcommand{\wrt}{w.r.t.\ }
\newcommand{\st}{s.t.\ }
\newcommand{\newitem}{\begin{center}\rule{4in}{1pt}\end{center}}
\newcommand{\calB}{\mathcal{B}}
\newcommand{\bhat}{\hat{b}}
\newcommand{\indicator}{\mathbf{1}}
\newcommand{\xor}{\oplus}
\newcommand{\quadtextquad}[1]{\quad\text{#1}\quad}
\newcommand{\bid}{\mathtt{bid}}
\newcommand{\bminusi}{b_{-i}}
\newcommand{\bi}{b_{i}}
\newcommand{\bigpar}{\bigskip\par}

\newcommand{\clicks}{\mathcal{C}}
\newcommand{\price}{\mathcal{P}}

% some commands (yogi)
\theoremstyle{remark}
\newtheorem*{remark}{Remark}
\newcommand{\bbar}{\bar{b}}
\newcommand{\yogicomment}[1]{\marginpar{\scriptsize\sl\raggedright YS: #1}}
\newcommand{\psim}{\ensuremath{\mbox{\sc PSim}}}
\newcommand{\bex}{\ensuremath{\mathtt{BEX}}}
\newcommand{\B}{\mathcal{B}}
\newcommand{\frakS}{\mathfrak{S}}
\newcommand{\bfr}{\mathbf{r}}
\newcommand{\bfrhat}{\mathbf{\widehat{r}}}
\newcommand{\bfe}{\mathbf{e}}

\setlength{\fboxsep}{1.5pt}
\newcommand{\InFigureLAtRoundT}{\fbox{1}}
\newcommand{\InFigureIPrimeAtRoundTForLargeBid}{\fbox{2}}
\newcommand{\InFigureLDecreasesHerBid}{\fbox{3}}
\newcommand{\InFigureIIncreasesHerBid}{\InFigureLDecreasesHerBid}
\newcommand{\InFigureRoundTIsInfluential}{\fbox{4}}
\newcommand{\InFigureLIsJOrJPrime}{\fbox{5}}
\newcommand{\InFigureBIPlusRho}{\fbox{6}}
\newcommand{\InFigureBIPlusRhoPrime}{\fbox{7}}
\newcommand{\InFigureBIPlusRhoEqualsBIPlusRhoPrime}{\fbox{8}}

\theoremstyle{plain}
\newtheorem{observation}[theorem]{Observation}

\newpage

%%%%%%%%%%%%%%%%%%%%%%%%%%%%%%%%
\section{Introduction}
\label{sec:intro}

%AS1: Moshe was worried that this para is excessive for EC.
% Still, I think it is nice to have it for the sake of completeness.
% Also, even EC papers should be readable by non-EC audience...
In recent years there has been much interest in understanding the implication of strategic behavior on the performance of algorithms whose input is distributed among selfish agents. This study was mainly motivated by the Internet, the main arena of large scale interaction of agents with conflicting goals. The field of Algorithmic Mechanism Design~\cite{NR01} studies the design of mechanisms in computational settings (for background see the recent book~\cite{NRTV07} and survey~\cite{Rou08}).

Much attention has been drawn to the market for sponsored search (e.g.~\cite{LPSV07,EOS07,Varian07, MSVV07,Aggarwal-Muthukrishnan-focs08-tutorial}), a multi-billion dollar market with numerous auctions running every second. Research on sponsored search mostly focus on equilibria of the Generalized Second Price (GSP) auction~\cite{EOS07,Varian07}, the auction that is most commonly used in practice (e.g. by Google and Bing), or on the design of truthful auctions~\cite{AGM06}. All these auctions rely on knowing the rates at which users click on the different advertisements (a.k.a. click-through rates, or CTRs), and do not consider the process in which these CTRs are learned or refined over time by observing users' behavior. We argue that strategic agents would take this process into account, as it influences their utility. While prior work~\cite{IJMT05} focused on the influence of click fraud on methods for learning CTRs, we are interested in the implications of the \emph{strategic bidding} by the agents. Thus, we consider the problem of designing truthful sponsored search auctions when the process of learning the CTRs is a part of the game.
% AS1: removed sentence -- we discuss it two pars later
% Our goal is to maximize social welfare.

We are mainly interested in the interplay between the online learning and the strategic bidding. To isolate this issue, we consider the following setting, which is a natural \emph{strategic} version of the multi-armed bandit (MAB) problem. In this setting, there are $k\geq 2$ agents. Each agent $i$ has a single advertisement, and a \emph{private} value $v_i> 0$ for every click she gets.  The mechanism is an online algorithm that first solicits bids from the agents, and then runs for $T$ rounds. In each round the mechanism picks an agent (using the bids and the clicks observed in the past rounds), displays her advertisement, and receives a feedback -- if there was a click or not. Payments are charged after round $T$. Each agent tries to maximize her own utility: the value that she derives from clicks minus the payment she pays. We assume that initially no information is known about the likelihood of each agent to be clicked, and in particular there are no Bayesian priors.

We are interested in designing mechanisms which are truthful (in dominant strategies): every agent maximizes her utility by bidding truthfully, for any bids of the others and {\em for any clicks} that would have been received
% MB5.8.2013: added
(that is, for any realization of the clicks an agent never regrets being truthful in retrospect).
%AS1: here is a less adversarial note
The goal is to maximize the social welfare.%
\footnote{Social welfare includes both the auctioneer's revenue and the agents' utility. Since in practice different sponsored search platforms compete against one another, taking into account the agents' utility increases the platform's attractiveness to the advertisers.}
%
% MB2.5: this is to fight the crap about revenue and to enlighten the reader about the problem in the other paper. We might like to move to the body of the paper.
\OMIT{\footnote{An alternative goal is to maximize the revenue, but there are several reasons we focus on social welfare and not revenue.
First, the goal of maximizing revenue (implicitly) assumes that the auctioneer is a monopolist, cares only about his own payoff and does not care at all about the welfare of the advertisers. In sponsored search there is clearly competition between search platforms and by maximizing welfare and not revenue a platform increases its attractiveness to the advertisers (additionally, its is unrealistic to assume the auctioneer can extract monopolist rents when facing competition).
Secondly, welfare loss correspond directly to regret, which is well defined even for a non-truthful algorithm (with respect to its input). Thus we can directly compare the quality of truthful mechanisms to unrestricted algorithms it terms of regret.
On the other hand, the regret of a non-truthful algorithm does not seem to relate to revenue in a natural way, thus
% there is no natural definition of revenue for algorithms that are not truthful, so
it is unclear how one can one use regret to compare truthful mechanisms to unrestricted algorithms in terms of revenue.
}}%
Since the payments cancel out, this is equivalent to maximizing the total value derived from clicks, where an agent's contribution to that total is her private value times the number of clicks she receives.
% The goal is to maximize the social welfare. Since the payments cancel out, this is equivalent to maximizing the value derived from clicks. Thus, the ``quality" of an agent is her private value times the number of clicks she receives. Social welfare is maximized if the ``best" agent is chosen in each round.
We call this setting the \emph{\problem}.

%AS2 replaced this footnote with an inlined sentence.
\OMIT{\footnote{In the strategic version, the ``payoff" from choosing a given agent is her private value if the ad is clicked, and $0$ otherwise.}}
%AS2 added a footnote on welfare vs revenue
In the absence of strategic behavior this problem reduces to a standard MAB formulation in which an algorithm repeatedly chooses one of the $k$ alternatives (``arms'') and observes the associated payoff: the value-per-click of the corresponding ad if the ad is clicked, and $0$ otherwise. The crucial aspect in MAB problems is the tradeoff between acquiring more information (\emph{exploration}) and using the current information to choose a good agent (\emph{exploitation}). MAB problems have been studied intensively for the past three decades. In particular, the above formulation is well-understood~\cite{bandits-ucb1,bandits-exp3,Hayes-soda06} in terms of \emph{regret} relative to the benchmark which always chooses the same ``best'' alternative (\emph{time-invariant benchmark}).
This notion of regret naturally extends to the strategic setting outlined above, the total payoff being exactly equal to the social welfare, and the regret being exactly the loss in social welfare relative to the time-invariant benchmark.
%AS2: a slight change here
% \OMIT{providing a concrete connection between MAB algorithms and MAB mechanisms.}
Thus one can directly compare MAB algorithms and MAB mechanisms in terms of welfare loss (regret).
% MB2: I do not understand the sentence below
% Note that for such quantitative comparison on needs to maximize social welfare rather than revenue or some other objective.

Broadly, we ask how the design of MAB algorithms is affected by the restriction of truthfulness: what is the difference between the best {\em algorithms} and the best {\em truthful mechanisms}? We are interested both in terms of the structural properties and the gap in performance (in terms of regret). In short, we establish that the additional constraints imposed by truthfulness severely limit the structure and performance of online learning algorithms. We are not aware of any prior work that characterizes truthful online learning algorithms or proves negative results on their performance.

\xhdr{Discussion.}
We believe that the fundamental limitations of truthfulness are best studied in simple models such as the one defined above. We did not attempt to incorporate many additional aspects of pay-per-click ad auctions such as information that is revealed to and by agents over time, multiple ad slots, user contexts, ad features, etc. However, intuition from our impossibility results applies to richer models, and for some of these models it is not difficult to produce precise corollaries. The key idea in the simple truthful mechanism that we present (separating exploration and exploitation) can be easily extended as well.

We consider a strong notion of truthfulness: bidding truthfully is optimal for \emph{every} possible \realization (and bids of others). This notion is attractive as it does not require the agents to be risk neutral with respect to the randomness inherent in clicks, or consider their beliefs about the CTRs. It allows for the CTRs to change over time, and still incentivizes agents to be truthful. Moreover, an agent never regrets truthful bidding in retrospect. It is desirable to understand what can be achieved with this notion before moving to weaker notions, and thus we focus on this notion in this paper.

% ------------------------------
% [MOSHE: stopped editing here]

%  Another paper~\cite{BLNS08} has recently showed that there can be a large communication overhead in computing payments (over just the allocation) that are needed for truthful implementation.

% Scheduling unrelated machines Nisan Ronen: lower bound 2 (truthful), Algorithmic: prefect solution (2 machines, 3 jobs)
% Archer Tardos (Single Parameter) weighted sum of completion time:
% Lavi Nisan online (n vs. constant)
% Computational : Lavi Mu'alem Nisan. Mukund Shahar.

% \cite{PO06} modeled advertisement placement with the goal of revenue maximization in a Pay-per-Click auction as a multi-armed bandit problem.
% Maybe the closest paper to ours is a recent paper on the use of learning (multi-arm bandit algorithms) for Cost-Per-Action sponsored search auctions~\cite{NSV08}.[MOSHE: EXPAND!!!!]

% Other papers study other issues in sponsored search auctions, such as the online problem of matching queries to advertisements~\cite{MSVV07}.

% [MOSHE: still need to complete this part, probably by merging in some of the "related work" section]

\subsection{Our contributions}
\label{sec:contributions}

We present two main contributions: structural characterizations of (dominant-strategy) deterministic truthful mechanisms, and lower bounds on the regret that such mechanisms must suffer. The regret suffered by truthful mechanisms is significantly larger than the regret of the best MAB algorithms. We emphasize that our characterization results hold regardless of whether the mechanism's goal is to maximize welfare, revenue, or any other objective.

Formally, a mechanism for the \problem\ is a pair $(\A,\mathcal{P})$, where $\A$ is the \emph{allocation rule} (essentially, an MAB algorithm which also gets the bids as input), and $\mathcal{P}$ is the \emph{payment rule} that determines how much to charge each agent. Both rules can depend only on the observable quantities: submitted bids and click events (clicks or non-clicks) for ads that have been displayed by the algorithm. Since the allocation rule is an online algorithm, its decision in a given round can only depend on the click events observed in the past.

The distinction between an allocation rule and a payment rule is essential in prior work on Mechanism Design, and it is also essential for this paper. In particular, social welfare (and therefore regret) is completely determined by the allocation rule. This is because welfare includes each payment twice, with opposite signs: amount paid by an advertiser and amount received by the mechanism, and the two cancel out.

\xhdr{Characterization.}
The MAB mechanisms setting is a \emph{single-parameter auction}, the most studied and well-understood type of auctions. For such settings truthful mechanisms are fully characterized~\cite{Myerson, ArcherTardos}: a mechanism is truthful if and only if the allocation rule is monotone (by increasing her bid an agent cannot cause a decrease in the number of clicks she gets), and the payment rule is defined in a specific and, essentially, unique way. Yet, we observe that this characterization is {\em not} the right characterization for the MAB setting! The main problem is that if an agent is not chosen in a given round then the corresponding click event is not observed by the mechanism, in the sense that the mechanism does not know whether this agent would have received a click had it been selected in this round. Therefore the payment cannot depend on any such unobserved click events. This is a non-trivial restriction because the naive payment computation according to the formula mandated by~\cite{Myerson,ArcherTardos} requires simulating the run of the allocation rule for bids different than the ones actually submitted, which in turn may depend on unobserved click events.  We show that this restriction has severe implications on the structure of truthful mechanisms.

The first notable \emph{necessary} property of a truthful MAB mechanism is a much stronger version of monotonicity which we call ``pointwise monotonicity'':

\begin{definition}
\label{def:pwm}
A {\em \realization} consists of click information for all agents and all rounds: it specifies whether a given agent receives a click if it is selected in a given round.%
\footnote{Note that an MAB mechanism does not observe the entire \realization: it only observes click information for one agent per round, the agent that was selected in this round.}
An allocation rule is \emph{pointwise monotone} if for each \realization, each bid profile and each round, if an agent is selected at this round, then she is also selected after increasing her bid (fixing everything else).
\end{definition}

We first consider the case of two agents and
show that truthful MAB mechanisms must have a strict separation between exploration and exploitation, in the following sense. A crucial feature of exploration is the ability to influence the allocation in forthcoming rounds. To make this point more concrete, we call a round $t$ \emph{influential} for a given \realization, with influenced agent $j$, if for some bid profile changing the \realization for this round can affect the allocation of agent $j$ in some future round. We show that in any influential round, the allocation can not depend on the bids. Thus, we show that influential rounds are essentially useless for exploitation.

\begin{definition}
\label{def:exploration-separated}
An MAB allocation rule \A\ is called \emph{exploration-separated} if for any \realization,
the allocation in any influential round does not depend on the bids.
\end{definition}

In our model, agents derive value from clicks. In particular, an agent with zero value per click receives no value. We focus on mechanisms in which a truthfully bidding agent with zero value-per-click pays exactly zero; we call such mechanisms \emph{normalized}.  Among truthful single-parameter mechanisms, normalized mechanisms are precisely the ones that satisfy two desirable properties: \emph{voluntary participation} (truthfully bidding agents never lose from participating), and \emph{no positive transfers} (advertisers are charged, not paid).

% OLD:
% As is standard in the literature on Mechanism Design, we focus on mechanisms in which each agent's payment (averaged over clicks) is between $0$ and her bid; such mechanisms are called \emph{normalized}. These mechanisms are precisely the ones that satisfy two desirable properties: voluntary participation (truthfully bidding agents never lose from participating), and no positive transfers (advertisers are charged, not paid).

We also make a mild assumption that an allocation rule is \emph{scale-free}: invariant under multiplying all bids by the same positive number, i.e. does not depend on the choice of the currency unit. Many MAB algorithms from prior work can be easily converted into scale-free MAB allocation rules via some generic ways to incorporate bids into algorithms' specification.%
\footnote{Many algorithms from prior work on stochastic MAB maintain an estimate $\nu_i$ of the expected reward for each arm $i$, such as an upper confidence bound in $\UCB$~\cite{bandits-ucb1} or an independent sample from Bayesian posterior in Thompson's Heuristic~\cite{Thompson-1933}, so that the algorithms' decisions depend only on these estimates. An allocation rule can interpret $\nu_i$ as an estimate of the CTR, and use $\nu'_i = b_i\, \nu_i$ instead of $\nu_i$ for all decisions. Moreover, any MAB algorithm can be converted to a scale-free MAB allocation rule by assigning a reward of $b_i/ (\max_j b_j)$ to each agent $i$ for each click on her ad. We use both approaches in this paper, in Section~\ref{sec:naive} and Section~\ref{app:PSim}, respectively.}

We are now ready to present our main structural result for two agents.

\begin{theorem} \label{thm:main-characterization-2-agents}
Consider the \problem with two agents. Let \A\ be a non-degenerate,%
\footnote{Non-degeneracy is a mild technical assumption, formally defined in ``preliminaries'', which ensures that (essentially) if a given allocation happens for some bid profile  $(b_i, b_{-i})$ then the same allocation happens for all bid profiles $(x, b_{-i})$, where $x$ ranges over some non-degenerate interval. Without this assumption, all structural results hold (essentially) \emph{almost surely} w.r.t the $k$-dimensional Lebesgue measure on the bid vectors. Exposition becomes significantly more cumbersome, yet leads to the same lower bounds on regret. For clarity, we assume non-degeneracy throughout this paper.}
deterministic, scale-free allocation rule.
Then a mechanism $(\A,\mathcal{P})$ is normalized and truthful for some payment rule $\mathcal{P}$ if and only if \A\ is pointwise monotone and exploration-separated.
\end{theorem}

The case of more than two agents requires slightly more refined notions.
%We present a complete characterization that holds for any number of agents.

\begin{definition}
For a given realization and bid profile, a round is {\em secured} from an agent if that agent cannot change the allocation at that round by increasing his bid.
A deterministic MAB allocation rule is called \emph{weakly separated} if for every \realization and bid profile, if a round is influential for this realization and bid profile, then it is secured from every agent that this round influences.
\end{definition}

The ``weakly separated'' condition is weaker than ``exploration-separated'': while the latter ensures that all agents cannot change the allocation at any given influential round $t$, the former only requires this for each agent that is influenced by round $t$, fixing the bids of all other agents.
For two agents and a scale-free MAB allocation rule, the two conditions are equivalent.

Our complete characterization for any number of agents follows.

\begin{theorem} \label{thm:main-characterization}
Consider the \problem. Let \A\ be a non-degenerate deterministic allocation rule.
Then a mechanism $(\A,\mathcal{P})$ is normalized and truthful for some payment rule $\mathcal{P}$ if and only if $\A$ is pointwise monotone and weakly separated.
\end{theorem}

Note that the general characterization does not require the allocation rule to be scale-free. In the special case of two agents and scale-free allocation rules it implies Theorem~\ref{thm:main-characterization-2-agents}.

We also investigate under which assumptions a weakly separated MAB allocation rule is exploration-separated, as the latter condition is sufficient for proving performance limitations (bounds on regret). To this end, we adapt a well-known notion from the literature on Social Choice, called {\em Independence of Irrelevant Alternatives} (\emph{IIA}, for short): an MAB allocation rule is \emph{IIA} if for any given \realization, bid profile and round, a change of bid of agent $i$ cannot transfer the allocation in this round from agent $j$ to agent $l$, where these are three distinct agents.
Note that the IIA condition trivially holds if there are only two agents. We prove that for
a non-degenerate deterministic allocation rule which is scalefree, pointwise monotone, and satisfies IIA it holds that the rule is  exploration-separated if and only if it is weakly separated.
Technically, assuming IIA allows us to extend our performance limitations results to more than two agents.%
\footnote{%
Since prior work on MAB algorithms did not address strategic issues,  these algorithms were not designed to satisfy properties like (pointwise) monotonicity and IIA (and besides, these properties are not even well-defined for MAB \emph{algorithms}, only for MAB \emph{allocation rules}). So it is not yet clear how limiting are these properties. The simple pointwise monotone MAB allocation rule described later in the Introduction does satisfy IIA, but suffers from high regret. Designing better-performing MAB allocation rules that are (pointwise) monotone appears quite challenging. For instance, such allocation rule is one of the main results in the follow-up paper~\cite{Transform-ec10}. We leave open the question of existence of low-regret MAB allocation rules that are both pointwise-monotone and IIA.}

% OLD by Alex:
% Since prior work on MAB algorithms did not address the strategic issues such as (pointwise) monotonicity, it is not clear how limiting is the additional assumption of IIA. The simple (and inefficient) pointwise monotone MAB allocation rule described later in the Introduction does satisfy IIA. Designing more efficient MAB allocation rules with (pointwise) monotonicity appears quite challenging. For instance, such allocation rule is one of the main results in the follow up paper~\cite{Transform-ec10}.

\OMIT{ % Moshe's version
Every allocation rule that is exploration-separated is also weakly separated, but not vise-versa. Yet, we are able to present sufficient conditions on the allocation rule that insure that the two are equivalent. An allocation rule is \emph{scale-free} if it is invariant under multiplying all bids by the same positive number (essentially, changing the currency unit).
An allocation rule is {\em Independent of Irrelevant Alternatives} (\emph{IIA}, for short) if for any given \realization, bid profile and round, a change of bid of agent $i$ cannot transfer the allocation in this round from agent $j$ to agent $l$, where these are three distinct agents.
Note that the IIA condition trivially holds if there are only two agents. We prove that for
a non-degenerate deterministic allocation rule which is scalefree, pointwise monotone, and satisfies IIA it holds that the rule is  exploration-separated if and only if it is weakly separated. %Thus any lower bounds we prove on regret for
%MB5.9.2013: need to decide if we want to make the above a formal claim.
}

\xhdr{Lower bounds on regret.}
In view of the characterizations of truthful mechanisms, %Theorem~\ref{thm:main-characterization},
we present a lower bound on the performance of exploration-separated algorithms. We consider a setting, termed the \emph{stochastic \problem}, in which each click on a given advertisement is an independent random event which happens with a fixed probability, a.k.a. the CTR. The expected ``payoff'' from choosing a given agent is her private value times her CTR. For the ease of exposition, assume that the bids lie in the interval $[0,1]$. Then the non-strategic version is the \emph{stochastic MAB problem} in which the payoff from choosing a given arm $i$ is an independent sample in $[0,1]$ with a fixed mean $\mu_i$. In both versions, we compete with the \emph{best-fixed-arm benchmark}: the hypothetical allocation rule (resp.\ algorithm) that always chooses an arm with the maximal expected payoff. This benchmark is standard in the literature on stochastic MAB; it is optimal among all MAB algorithms that are given the expected rewards for each arms (resp., among all MAB allocation rules that are given the bids and the CTRs).
We define \emph{regret} as the expected difference between the social welfare
(resp.\ total payoff) of the benchmark and that of the allocation rule (resp.\ algorithm). The algorithm's goal is to minimize $R(T)$, worst-case regret over all problem instances on $T$ rounds.

\OMIT{%%%%%
In the stochastic \problem\ for each agent clicks are distributed IID with some expectation that is not known to the algorithm and the goal is to minimize $R(T)$, the worst-case expected regret over all instances. We obtain lower bounds on regret, the difference in social welfare between the algorithm and the benchmark which always selects the same agent.
}%%%%%%%%

We show that the worst-case regret of any exploration-separated algorithm is \emph{larger} than that of the optimal MAB algorithm~\cite{bandits-exp3}: $\Omega(T^{2/3})$ vs. $O(\sqrt{T})$ for a fixed number of agents.
We obtain an even more pronounced difference if we restrict our attention to the \emph{$\delta$-gap} problem instances: instances for which the best agent is better than the second-best by a (comparatively large) amount $\delta$, that is $\mu_1 v_1 - \mu_2 v_2 = \delta \cdot (\max_{i}v_i)$, where arms are arranged such that $\mu_1 v_1 \geq \mu_2 v_2 \geq \cdots \geq \mu_k v_k$. Such problem instances are known to be easy for the MAB algorithms. Namely, an MAB algorithm can concurrently achieve the optimal worst-case regret $O(\sqrt{kT \log T})$ \emph{and} regret $O(\tfrac{k}{\delta}\,\log T)$ on $\delta$-gap instances~\cite{Lai-Robbins-85,bandits-ucb1}. However, we show that for exploration-separated allocation algorithms the worst-case regret $R_{\delta}(T)$ over the $\delta$-gap instances is polynomial in $T$ (rather than poly-logarithmic in $T$) as long as worst-case regret is even remotely non-trivial (i.e., sublinear). Thus, for the $\delta$-gap instances the gap in the worst-case regret between unrestricted algorithms and exploration-separated algorithms is {\em exponential} in $T$.

\begin{theorem}\label{thm:LB-k}
Consider the stochastic \problem\ with $k\geq 2$ agents. Let \A\ be a deterministic allocation rule that is exploration-separated. Then \A\ has worst-case regret
    $R(T) = \Omega(k^{1/3}\, T^{2/3})$.
Moreover,
if $R(T)=O(T^\gamma)$ for some $\gamma<1$ then for every fixed
    $\delta\leq \tfrac14$ and any $\eps>0$
the worst-case regret over the $\delta$-gap instances is
    $R_\delta(T) = \Omega(\delta\,T^{2(1-\gamma)-\eps})$.
\end{theorem}

% \mbedit{Given our characterization (Theorem~\ref{thm:main-characterization-2-agents})},
For two agents, Theorem~\ref{thm:LB-k} implies a significant gap in performance between truthful MAB mechanisms and the best MAB algorithms, since truthful MAB mechanisms are necessarily exploration-separated.%
\footnote{Formally, this holds for truthful MAB allocation rules with allocation rules that satisfy the mild assumptions of non-degeneracy and scale-freeness. We remove the latter assumption in one of the extensions.}
For example, while truthful MAB mechanisms suffer regret of $\Omega(T^{2/3})$, the best algorithms have regret of only $O(\sqrt{T})$; as we described above, for $\delta$-gap distances the difference in regret is even more pronounced.

For more than two agents, Theorem~\ref{thm:LB-k} does not immediately imply any regret bounds for truthful MAB mechanisms. This is because the theorem requires the ``exploration-separated'' condition, whereas the corresponding characterization result in Theorem~\ref{thm:main-characterization} only guarantees the ``weakly separated'' condition. Recall that one way to guarantee the  ``exploration-separated'' condition (and therefore the regret bound) is to furthermore assume IIA. It is an open question whether one can prove similar regret bounds for weakly separated MAB allocation rules
without assuming IIA.

\OMIT{
while Theorem~\ref{thm:LB-k} requires the stronger ''exploration-separated'' condition.
The regret bounds \asdelete{will} hold for any truthful \asedit{MAB} mechanisms with more than two agents as long as we can show that their allocation \asedit{rule is exploration-separated, which} is indeed the case for allocation rules that satisfy IIA.
\mbdelete{We recover the regret bounds in Theorem~\ref{thm:LB-k} if we further assume that the allocation rule satisfies IIA.}
}

% We prove that for a non-degenerate deterministic allocation rule which is scalefree, pointwise monotone, and satisfies IIA it holds that the rule is  exploration-separated if and only if it is weakly separated.

\OMIT{ %%% Moshe's version
Since for two agents truthful mechanisms are necessarily exploration-separated by Theorem~\ref{thm:main-characterization-2-agents},
Theorem~\ref{thm:LB-k} implies that the gaps in performance between exploration-separated algorithms and the best algorithms translates to
gaps in performance between truthful mechanisms and the best algorithms.
For more than two agents the lower bounds of Theorem~\ref{thm:LB-k} do not immediately imply lower bounds on the performance of truthful mechanisms as (by Theorem~\ref{thm:main-characterization}) a necessary condition for a mechanism to be truthful in this case is for the allocation to be weakly-separated and not necessarily satisfy the stronger condition of exploration-separated.
For the case that the allocation is scale-free and satisfies IIA we have shown that these two conditions are equivalent and thus the lower bounds hold. One of the open questions we leave is proving lower bounds on weakly-separated allocations (which will imply lower bounds on truthful mechanisms for more than two agents).
} %%%%%%%%

\OMIT{ %%%%%%%%%%
That is, if the algorithm has weakly sublinear worst-case regret (regret of $O(T^\gamma)$ for some $\gamma<1$) it must suffer high worst-case expected regret of $R_\delta(T) = \Omega(\delta\, T^{1-\gamma})$ on $\delta$-separated instances.
This in contrast to non-truthful algorithms:
there exists an algorithm~\cite{bandits-ucb1} with $R(T)=O(\sqrt{T})$ and
$R_{\delta}(T) = O(\log T/{\delta})$ for any $\delta>0$.
% \MBnote{Can we replace $O_{\delta}(\log T)$ by $O(\delta\cdot \log T)$?}
Thus if we require weakly sublinear worst-case regret
then there is an \emph{exponential} gap between the best
worst-case expected regret for $\delta$-separated instances of algorithms and of truthful mechanisms!
} %%%%%%%

We note that our lower bounds hold for a more general setting in which the values-per-click can change over time, and the advertisers are allowed to change their bids at every time step.

Somewhat counter-intuitively, the lower bound on regret for $k=2$ agents does not immediately imply the same lower bound for any constant $k>2$. This is, essentially, because our setting requires a mechanism to show an ad in each round. A seemingly obvious approach to extend the lower bound from $k=2$ to (say) $k=3$ is to assume, for the sake of contradiction, that there exists a truthful MAB mechanism $\calM$ for $3$ agents whose regret is less than the lower bound for two agents, and use $\calM$ construct a truthful MAB mechanism $\calM'$ for two agents with the same regret. (This would yield a contradiction, and hence prove the lower bound for three agents.) The derived two-agent mechanism $\calM'$ adds a fictitious third agent (a dummy) that never receives any clicks, and runs the original three-agent mechanism $\calM$. However, when $\calM$ picks the dummy agent, the two-agent mechanism must pick one of the two real agents. These additional allocations may distort the agents' incentives, so $\calM'$ is not guaranteed to be truthful. Hence, this reduction is not guaranteed to work. Likewise, the allocation rule of $\calM'$ is not guaranteed to be weakly separated even if the allocation rule of $\calM$ is exploration-separated. Thus, we cannot immediately obtain a lower bound on regret for more than two agents simply by combining the two-agent characterization in Theorem~\ref{thm:main-characterization-2-agents} and the two-agent regret bound of Theorem~\ref{thm:LB-k}.

\xhdr{Tightness: a positive result.}
To complete the picture for exploration-separated MAB allocation rules,
we present a very simple deterministic mechanism that is truthful and normalized, and matches the lower bound
    $ R(T) = \Omega(k^{1/3}\, T^{2/3})$
up to logarithmic factors. The allocation rule in this mechanism is exploration-separated; it consists of two phases: an exploration phase in which agents are chosen in a round-robin fashion, followed by an exploitation phase which allocates all rounds to the agent with the best empirical performance in the exploration phase. Crucially, the duration of the exploration phase is fixed in advance (and optimized given $k$ and $T$).

\xhdr{Extensions.}
We extend our main results in several directions.

\begin{enumerate}

\item
% \fakeItem[1.]
We derive a lower bound on regret for deterministic truthful mechanisms without assuming that the allocations are scale-free. In particular, for two agents there are no assumptions. This lower bound holds for any $k$ (the number of agents) assuming IIA. However, the value of the lower bound does not increase with $k$; in this sense this lower bound is weaker than the one in Theorem~\ref{thm:LB-k}.

%%%%%%%%%%%%%%%%%%
\item
%\fakeItem[2.]
We consider randomized MAB mechanisms that are \emph{universally truthful}, i.e. truthful for each realization of the internal random seed. We extend the $\Omega(k^{1/3}\,T^{2/3})$ lower bounds on regret to mechanisms that randomize over exploration-separated deterministic MAB allocation rules.

\OMIT{For mechanisms that randomize over exploration-separated deterministic allocation rules, we obtain the same lower bounds as in Theorem~\ref{thm:LB-k}.}

%%%%%%%%%%%%%%%%%%
\item
%\fakeItem[3.]
We consider randomized MAB mechanisms under a weaker (less restrictive) version of truthfulness: a mechanism is \emph{weakly truthful} if for each \realization, it is truthful in expectation over its random seed. We show that any randomized allocation that is pointwise monotone and satisfies a certain stong notion of ``separation between exploration and exploitation'' can be turned into a mechanism that is weakly truthful and normalized.

We apply this result to the version of the \problem\ in which the clicks are chosen by an oblivious adversary.\footnote{An oblivious adversary chooses the entire \realization in advance, without observing algorithm's behavior.} (The corresponding algorithmic version is the \emph{adversarial MAB problem}~\cite{bandits-exp3,CesaBL-book}.) Using an MAB algorithm from the literature~\cite{Bobby-stoc04-journal-version,Robert-Kleinberg-Lecture-8}, we obtain a weakly truthful MAB mechanism for this problem with regret $\bigo( (k\log k)^{1/3}\cdot T^{2/3})$. This matches our lower bound for deterministic MAB mechanisms up to $(\log k)^{1/3}$ factor.

%~\cite{bandits-exp3,Hayes-soda06,Abernethy-Hazan-Rakhlin-08-Competing,Bartlett-etal-colt08-High-probability}.)

%%%%%%%%%%%
\item
%\fakeItem[4.]
The stochastic \problem\ admits a very reasonable notion of truthfulness that is even weaker: \emph{truthfulness in expectation}, where for each vector of CTRs the expectation is taken over clicks (and the internal randomness in the mechanism, if the latter is not deterministic).\footnote{\emph{Normalized-in-expectation} and \emph{monotone-in-expectation} properties are defined similarly.} Following our line of investigation, we ask whether restricting a mechanism to be truthful in expectation has any implications on the structure and regret thereof.
    Given our negative results on mechanisms that are truthful and normalized, it is tempting to seek similar results for mechanisms that are truthful in expectation and normalized in expectation. We show that such approach is not likely to be fruitful.

    Surprisingly, we prove that any monotone-in-expectation MAB allocation rule gives rise to an MAB mechanism that is truthful in expectation and normalized in expectation, with a very minor increase in regret. The key idea is to view the expected payments as multivariate polynomials over the CTRs, and argue that any such polynomial can be ``implemented'' by a suitable payment rule. While this result is purely theoretical, e.g. because the payments have very high variance, it implies that any impossibility result for truthful-in-expectation MAB mechanisms must either follow directly from monotonicity-in-expectation of the allocation rule, or requires bounds on the variability of the payments.

    \OMIT{\footnote{The follow-up work~\cite{Transform-ec10} rules out the former possibility by showing that an optimal MAB algorithm gives rise to a monotone-in-expectation MAB allocation rule.}}
\end{enumerate}

\OMIT{We also provide a number of extensions.
First, we prove a similar (but slightly weaker) regret bound without the scale-free assumption. Second, we extend some of our results to randomized mechanisms; in this setting, (dominant-strategy) truthfulness means ``truthfulness for each \realization of the private randomness''. Third, we consider a weaker notion of truthfulness for randomized mechanisms -- for each \realization of the clicks, but in expectation over the random seed, and use this notion to provide algorithmic results for the version of the \problem\ in which clicks are chosen by an adversary. Fourth, we discuss an even more permissive notion of truthfulness -- truthfulness in expectation over the clicks (and the random seed).}

\OMIT{
On a high level, we have defined a simple model to study the interplay between online learning and mechanism design, and more specifically the limitations imposed by truthfulness. We have established that truthfulness requirement can severely limit the structure and performance of online learning algorithms. To the best of our knowledge, this is the first such result.}

\xhdr{Informational obstacle.}
Our paper exposes a new kind of obstacle which might stands in the way of designing truthful mechanisms: insufficient observable information to compute payments; we will term it ``informational obstacle'' from here on.

Interestingly, this obstacle appears more general than the current setting. First, it would still feature prominently in any mechanism design setting which can be modeled as one of the numerous MAB settings studied in the literature. Second, and perhaps more importantly, we conjecture that it can be extended to a very general class of mechanisms that interact with the environment. The follow-up work~\cite{SingleCall-ec12,Parkes-netecon12} provides some evidence to this conjecture, see Section~\ref{sec:followup} for more details.
% OLD by Alex: Interestingly, this obstacle appears more general than the current setting. First, it would still feature prominently if we plug in any one of the numerous MAB settings studied in the literature. Second, and perhaps more importantly, we conjecture that it can be extended to a very general class of mechanisms that interact with the environment. The follow-up work~\cite{SingleCall-ec12,Parkes-netecon12} provides some evidence to this conjecture, see Section~\ref{sec:followup} for more details.

\OMIT{
This paper prompts a number of open questions, some of which have been resolved in the follow-up work~\cite{Transform-ec10,SingleCall-ec12,Gatti-ec12}. This follow-up work is detailed in Section~\ref{sec:followup}, and the current snapshot of open questions is in Section~\ref{sec:questions}.}

\subsection{Additional related work}
\label{sec:related-work}

\xhdr{Mechanism Design.} The question of how the performance of a truthful mechanism compares to that of the optimal algorithm for the corresponding non-strategic problem is one of the central themes in Algorithmic Mechanism Design.
%considered in the literature in a number of other auction settings.
Performance gaps have been shown for various scheduling problems~\cite{ArcherTardos,NR01,DS08} and for online auction for expiring goods~\cite{LN05}. Other papers presented approximation gaps due to \emph{computational constraints}, e.g. for combinatorial auctions~\cite{LMN03, DS08} and combinatorial public projects~\cite{PSS08},
showing a gap via a structural result for truthful mechanisms.

The intersection of Machine Learning and Mechanism Design is an active research area which includes work in various topics such as
online mechanisms~\cite{LN05},
dynamic auctions~\cite{BV06,AS07},
dynamic pricing~\cite{Roths74},
secretary problems~\cite{Freeman-survey83},
offline learning from self-interested data sources~\cite{Balcan-jcss08,Meir-AI12}
and a number of others. A more detailed review of this area, or any of the topics listed above, is beyond the scope of this paper.

\OMIT{
Interestingly, MAB algorithms feature prominently in two other settings in Mechanism Design which are very different from MAB mechanisms studied in this paper. One is dynamic pricing, which reduces to a bandit-style problem in which ``arms'' correspond to different prices. The other setting concerns user-generated content, namely selecting among reviews provided by self-interested agents.}

\OMIT{The study of MAB mechanisms has been initiated by Gonen and Pavlov~\cite{GonenP07}. The authors present a MAB mechanism which is claimed to be truthful in a certain approximate sense. Unfortunately, this mechanism does not satisfy the claimed properties; this was also confirmed with the authors through personal communication (see also a similar note in~\cite{DevanurK09}).}

%AS1: added a para about Rica
\xhdr{MAB mechanisms.}
%AS1: rewording Moshe's paragraph.
%AS2: mentioned potential best-response dynamics
%MB2: changes below
MAB algorithms were used in the design of Cost-Per-Action sponsored search auctions in Nazerzadeh et al.~\cite{NSV08}, where the authors construct a mechanism with approximate (asymptotic) properties of truthfulness and individual rationality. However, even if the gains from lying are small,
% this solution concept is weaker than the exact notion and
it may still be rational for the agents to deviate from being truthful, perhaps significantly.
Moreover, as truthful bidding is not a Nash equilibrium, an agent may speculate that other agents will deviate, which in turn may increase her own incentives to deviate. All of that may result in unpredictable, and possibly highly suboptimal outcomes.
On the other hand, approximate truthfulness guarantees suffice whenever it is reasonable to assume that the agents would not lie unless it leads to significant gains.

% MB this sounds negative
% On the other hand, in the present

\OMIT{In this paper we focus on understanding what can be achieved with the \emph{exact} truthfulness, mainly proving results of structural and lower-bounding nature. We note in passing that providing similar results for the approximately truthful setting such as the one in~\cite{NSV08} is a worthy and challenging open question.}

%AS1: this is Moshe's paragraph about NSV08
\OMIT{While with the standard notion of exact truthfulness each agent maximizes her utility by being truthful, in the relaxed notions studied at~\cite{NSV08} it might still be completely rational for an agent to deviate from being truthful. If many agents deviate the outcome might be far from the desired one. Thus it seems desirable to understand what can be achieved with exact truthfulness and that is the focus of our paper.}

% MB2.5: We should consider citing Gonen and Pavlov and saying one of the following: 1) they are wrong, or 2) the try to give upper bounds, we focus on lower bounds...
\OMIT{\footnote{Another paper that aims to build MAB mechanisms with relaxed notion of truthfulness is~\cite{GonenP07}. Yet we have noticed  that the mechanism presented in~\cite{GonenP07} does not satisfy the claimed properties (as the price an agent pays depends on her bid), and this was confirmed with the authors through personal communication.}}
% MB2.5: we need to decide how to extend the above to stress our work is important even with NSV08 as we consider the more standard notion of truthfulness. We need to be careful not to annoy NSV though (Amin is on the PC).

In a concurrent and independent work with respect to this paper, Devanur and Kakade~\cite{DevanurK09} considered the same setting: deterministic truthful MAB mechanisms. They focus on maximizing the revenue of the mechanism (as opposed to the social welfare).
They present an impossibility result for the two-agent case: a lower bound of $\Omega(T^{2/3})$ on the loss in revenue with respect to the VCG payments; this bound is extended to deterministic MAB mechanisms that are truthful with high probability. They also provide a deterministic truthful mechanism which matches the above lower bound, and is almost identical to our simple two-phase mechanism described in Section~\ref{sec:contributions}.%
\footnote{This mechanism is for a more general setting in which values-per-click change over time and the agents are allowed to submit a different bid at every round. Instead of assigning all impressions to the same agent in the exploitation phase, their mechanism runs the same allocation and payment procedure for each exploitation round separately, with the bids submitted in this round.}

\OMIT{The results in~\cite{DevanurK09} contain  a structural characterization very similar to ours, but (to the best of our understanding) only for two agents and for a more general setting in which agents are allowed to bid in every round. (Note that truthfulness in this setting is more demanding.)}

%the one that we present in order to match
%the lower bound in Theorem~\ref{thm:LB-main}.)

% Hamid paper says about them: "Gonen and Pavlov [12] give a mechanism which learns the click-through rates via sampling and show that truthful bidding is, with high probability, a (weakly) dominant strategy in this mechanism.

A closely related line of work on \emph{dynamic auctions}~\cite{BV06,AS07,Segal-dynamic11,Kakade-pivot11} considers a more general setting in which private information is revealed to agents over time.
The mechanism needs to create the right incentives for the agents to reveal all the information they receive over time, and to stay in the auction after every round;
these challenges do not exist in our setting, in which all private information is known to the agents upfront.
% Such revelation of information over time makes satisfying the incentives and participation constraints a challenging task,  a challenge that does not exist in our setting in which all private information is known to the agents upfront.
On the other hand, these papers study fully Bayesian settings in which Bayesian priors on CTRs are known and VCG-like social welfare-maximizing mechanisms are therefore feasible. In our setting -- with no priors on CTRs -- VCG-style mechanisms cannot be applied as such mechanisms require the allocation to exactly maximize the expected social welfare, which is impossible (and even not well-defined) without a prior.
% OLD: In terms of pay-per-click ad auctions, this corresponds to multi-step auctions with repeated bids, \asedit{changing values-per-click}, and uncertainty over CTRs. \asdelete{and uncertainty on expected payoffs.} In addition to the significantly more complicated incentives structure \mbedit{(due to dynamic information)}\mbdelete{(due to repeated bids)},  a key technical distinction is that these papers study fully Bayesian settings in which Bayesian priors on CTRs are known and VCG-like social welfare-maximizing mechanisms are therefore feasible. In our setting -- with no priors on CTRs -- VCG-style mechanisms cannot be applied as such mechanisms require the allocation to exactly maximize the expected social welfare, which is impossible (and even not well-defined) without a prior.
Moreover, even if applied to MAB mechanisms with Baeysian priors over CTRs, the techniques from this line of work can only guarantee truthfulness in expectation over the Bayesian prior, which is a much weaker notion compared to the ``prior-independent" notions of truthfulness that are studied in this paper.

%Other papers study other issues in sponsored search auctions, such as the online problem of matching queries to advertisements~\cite{MSVV07}.

\xhdr{Multi-armed bandits (MAB).}
Absent the strategic constraint, our problem fits into the framework of MAB algorithms. MAB has a rich literature in Statistics, Operations Research, Computer Science and Economics; a reader can refer to~\cite{CesaBL-book,Bergemann-survey06} for background. Most relevant to the present paper is the work on stochastic MAB~\cite{Lai-Robbins-85,bandits-ucb1} and adversarial MAB~\cite{bandits-exp3}. Both directions have spawned vast amounts of follow-up research. Results used in this paper come from~\cite{bandits-ucb1,Lai-Robbins-85,bandits-exp3,Bubeck-colt09,Bobby-stoc04-journal-version,Robert-Kleinberg-Lecture-8}.

Our lower bounds on regret use (a novel application of) the relative entropy technique from~\cite{Lai-Robbins-85,bandits-exp3}, see~\cite{Bobby-LN} for an account. This is the technique typically used to prove lower bound on regret for MAB and related problems. For other application of this technique, see e.g.~\cite{Hayes-soda06,Bobby-Karp,LipschitzMAB-stoc08,Ben-Or-Hassidim-08-The-Bayesian}.

The prior work on MAB algorithms considered numerous MAB settings with various assumptions on payoff evolution over time (e.g.,~\cite{bandits-exp3,DynamicMAB-colt08,Hazan-soda09}), dependencies between arms (e.g.,~\cite{FlaxmanKM-soda05,yahoo-bandits-icml07,LipschitzMAB-stoc08,GPbandits-icml10}), side information available to an algorithm (e.g.,~\cite{LipschitzMAB-stoc08,Langford-nips07,contextualMAB-colt11}), etc. Many of these settings are motivated by pay-per-click ad auctions. For every such MAB setting one could define the corresponding version of the \problem.

\OMIT{ %%%%%%%%%
\footnote{Multi-arm bandit algorithms were used in the design of auctions for Cost-Per-Action sponsored search~\cite{NSV08}. While~\cite{NSV08} focuses on an upper bound with relaxed notions of truthfulness and individual rationality, we focus on lower bounds, the increase in regret due to exact truthfulness.}

(This is in contrast to the recent work on \emph{dynamic auctions}~\cite{BV06,AS07} which considers fully Bayesian settings in which there is a known prior on CTRs, and VCG-like social welfare-maximizing mechanisms are feasible.)

This paper focuses on understanding the gaps between performance of the best {\em algorithms} and the best {\em truthful mechanisms} for our problem (following question (Q1) of~\cite{Rou08}).
The general scheme for proving such a gap consists of two stages. First one characterizes the implementable allocation algorithms, and then one proves lower bounds on the best-achievable performance of an implementable algorithm.
% , without computational constraints.

Such a scheme was followed before in the literature.
A gap was proven between computationally feasible algorithms and truthful mechanism
in the approximation of combinatorial auctions~\cite{LMN03} and combinatorial public projects~\cite{PSS08}. Without computational constraints gap were shown for
online auction for expiring goods~\cite{LN05}, and for various scheduling problems~\cite{ArcherTardos,NR01}.
As far as we are aware of, no other paper before ours has used this general scheme to lower bound the regret of truthful multi-arm bandit algorithms.

} %%%%%%%%%%

\subsection{Follow-up work}
\label{sec:followup}

The conference publication of this paper gave rise to a several follow-up papers~\cite{Transform-ec10,SingleCall-ec12,Gatti-ec12,Parkes-netecon12} which have addressed some of the questions left open by this paper and posed some new ones. Below we present the current snapshot of this line of work.

One direction concerns weakly truthful, randomized MAB mechanisms. Informally, the main question here is whether they are significantly more powerful than their deterministic counterparts. Babaioff, Kleinberg and Slivkins~\cite{Transform-ec10} resolve this question in the affirmative: they prove that there exist weakly truthful randomized MAB mechanisms whose regret bounds for the stochastic MAB setting are optimal for MAB algorithms, both in the worst case and for $\delta$-gap instances. A major component of this result, henceforth called the \emph{BKS reduction}, reduces designing weakly truthful MAB mechanisms to designing MAB allocation rules that satisfy the appropriate notion of monotonicity called \emph{weak monotonicity}: an MAB allocation is \emph{weakly monotone} if for each \realization, it is monotone in expectation over its random seed.\footnote{\cite{Transform-ec10} uses a somewhat different (and perhaps more systematic) terminology regarding the different notions of truthfulness, monotonicity and normalization. We discuss the results from~\cite{Transform-ec10} using the terminology of the present paper.} The BKS reduction subsumes and generalizes our result on truthfulness in expectation (using a very different technique). Moreover, it is not specific to the stochastic MAB setting: it extends beyond MAB mechanisms to arbitrary \emph{single-parameter domains} (see~\cite{NRTV07} for more background). In particular, the BKS reduction applies to MAB mechanisms with clicks chosen by an oblivious adversary, and to MAB mechanism design problems based on most other settings studied in the vast literature on MAB algorithms.

\newcommand{\newUCB}{{\tt NewCB}\xspace}

Our truthful-in-expectation construction and the BKS reduction suffer from a very high variance in payments. Both results include an explicit tradeoff between the variance in payments and the loss in performance. Very recently, Wilkens and Sivan~\cite{SingleCall-ec12} have proved that the tradeoff in the BKS reduction is optimal in a certain \emph{worst-case} sense: the BKS reduction achieves the optimal worst-case variance in payments for any given worst-case loss in performance, where the worst case is over all monotone MAB allocation rules. (More generally, the optimality result in~\cite{SingleCall-ec12} applies to any given single-parameter problem.)

Additional developments in~\cite{Transform-ec10} concern MAB allocation rules. First, they prove that an MAB allocation rule based on \UCB{} satisfies monotonicity-in-expectation, and therefore can be transformed (using our result from Section~\ref{sec:expectation} or the BKS reduction) to a truthful-in-expectation MAB mechanism with essentially the same regret. Second, they provide a new deterministic MAB allocation rule called \newUCB which has optimal regret and is monotone. In conjunction with the BKS reduction, \newUCB yields the weakly truthful MAB mechanism discussed above.

The analysis in this paper provides a strong intuition that the crucial obstacle for deterministic MAB mechanisms is not the monotonicity of an allocation rule but instead the ``informational obstacle'': insufficient observable information to compute payments. The analysis of \newUCB in~\cite{Transform-ec10} makes this point rigorous. Moreover,~\cite{SingleCall-ec12,Parkes-netecon12} describe some additional settings, different from MAB mechanisms, where this ``informational obstacle'' arises. Wilkens and Sivan~\cite{SingleCall-ec12} provide two variants of  offline pay-per-click ad auctions with multiple ad slots. Shneider et al.~\cite{Parkes-netecon12} describe a packet scheduling problem in a network router, where the potentially non-observable information is the packet arrival times (rather than the click events). They observe that in the network router setting information about packet arrival times may be missing not only because it is not observed by the router but also because the router does not have much space to store it.

Finally, a very recent paper by Gatti, Lazaric and Trovo~\cite{Gatti-ec12} considers \emph{multi-slot MAB mechanisms}, i.e.  pay-per-click ad auctions with multiple ad slots and unknown CTRs. This setting combines multi-slot pay-per-click ad auctions~\cite{Varian07,EOS07} on the mechanism design side, and multi-slot MAB~\cite{RBA-icml08,Streeter08} on the learning side. The authors provide truthful multi-slot MAB mechanisms based on the simple MAB mechanism presented in this paper and (independently) in Devanur and Kakade~\cite{DevanurK09}.

Despite all these exciting development, MAB mechanisms are not well-understood; see Section~\ref{sec:questions} for the current snapshot of open questions.

\subsection{Map of the paper}

Section~\ref{sec:prelims} is preliminaries. Truthfulness characterization is developed and proved in Section~\ref{sec:structural} and Section~\ref{sec:IIA}. The lower bounds on regret are presented in Section~\ref{sec:regret}. The simple mechanism that matches these lower bounds is in Section~\ref{sec:naive}. Weakly truthful randomized allocations for adversarial clicks are derived in Section~\ref{sec:PSim}. Truthfulness in expectation is discussed in Section~\ref{sec:expectation}. Open questions are in Section~\ref{sec:questions}.

%Extensions and open questions are in Section~\ref{sec:extensions}.

%AS1: I'd like to mention appendices here...
% but I don't have a strong opinion on this
%To improve the flow of the paper, some of the material is moved to the appendices.
%Due to the page limit, many of the proofs are moved to the appendices.

%\input{sec-related-work} % Everything is commented in that file.

%%%%%%%%%%%%%%
\section{Definitions and preliminaries}
\label{sec:prelims}

In the \problem, there is a set $K$ of $k$ agents numbered from $1$ to $k$. Each agent $i$ has a \emph{value} $v_i>0$ for every click she gets; this value is known only to agent $i$.
Initially, each agent $i$  submits a \emph{bid} $b_i>0$, possibly different from $v_i$.
%MB2.5:
\footnote{One can also consider a more realistic and general model in which the value-per-click of an agent changes over time and the agents are allowed to change their bid at every round. The case that the value-per-click of each agent does not change over time is a special case. In that case truthfulness implies that each agent basically submits one bid as in our model (the same bid at every round), thus our main results (necessary conditions for truthfulness and regret lower bounds) also hold for the more general model.}
\footnote{Since private values $v_i$ are strictly positive, there is no need to allow zero bids. Also, this avoids some technical complications in the proofs. Accordingly, we define ``normalized mechanisms" in terms of the payment as $b_i\to 0$.}
The ``game'' lasts for $T$ rounds, where $T$ is the given \emph{time horizon}. A \emph{\realization} represents the click information for all agents and all rounds. Formally, it is a tuple
    $\rho=(\rho_1 \LDOTS \rho_k)$
such that for every agent $i$ and round $t$, the bit
    $\rho_i(t)\in \{0,1\}$
indicates whether $i$ gets a click if selected at round $t$. An \emph{instance} of the \problem\ consists of the number of agents $k$, time horizon $T$, a vector of private values
    $v = (v_1,\dots,v_k)$,
a vector of bids (\emph{bid profile})
    $b = (b_1,\dots,b_k)$,
and \realization $\rho$.

A \emph{mechanism} is a pair $(\A,\mathcal{P})$, where $\A$ is allocation rule and $\mathcal{P}$ is the payment rule. An \emph{allocation rule} is represented by a function $\A$ that maps bid profile $b$, \realization $\rho$ and a round $t$ to the agent $i$ that is chosen (receives an \emph{impression}) in this round: $\A(b;\rho;t) = i$. We also denote
    $\A_i(b;\rho;t)=\indicator_{\{\A(b;\rho;t) = i\}}$.
The allocation is \emph{online} in the sense that at each round it can only depend on clicks observed prior to that round. Moreover, it does not know the \realization in advance; in every round it only observes the \realization for the agent that is shown in that round. A \emph{payment rule} is a tuple
    $\mathcal{P}=(\mathcal{P}_1  \LDOTS \mathcal{P}_k)$,
where % for each agent $i$
    $\mathcal{P}_i(b;\rho)\in \mathbb{R}$
denotes the payment charged to agent $i$ when the bids are $b$ and the \realization is $\rho$.
% MB2.5:
\footnote{We allow the mechanism to determine the payments at the end of the $T$ rounds, and not after every round. This makes that task of designing a truthful mechanism {\em easier} and thus strengthen our necessary condition for truthfulness (the condition used to derive the lower bounds on regret.)}
Again, the payment can only depend on observed clicks.

A mechanism is called \emph{normalized} if for any agent $i$, bids $b_{-i}$ of the other agents, and \realization $\rho$ it holds that
    $ \calP_i(b_i,b_{-i};\, \rho) \to 0$ as $b_i\to 0$.
For any single-parameter, truthful mechanism, this limit exists and is independent of $b_i$~\cite{Myerson,ArcherTardos}; further, this limit is always $0$, for a given agent $i$, if and only if the payment per click is between $0$ and $b_i$.

For given \realization $\rho$ and bid profile $b$, the number of clicks received by agent $i$ is denoted
    $\mathcal{C}_i(b;\rho)$.
Call $\mathcal{C}=(\mathcal{C}_1  \LDOTS \mathcal{C}_k)$ the \emph{click-allocation} for \A. The \emph{utility} that agent $i$ with value $v_i$ gets from the mechanism $(\A, \mathcal{P})$ when the bids are $b$ and the \realization is $\rho$ is
    $\mathcal{U}_i(v_i;b;\rho) = v_i\cdot \mathcal{C}_i(b;\rho) - \mathcal{P}_i(b;\rho)$
(quasi-linear utility). The mechanism is \emph{truthful} if for any agent $i$, value $v_i$, bid profile $b$ and \realization $\rho$ it is the case that
    $\mathcal{U}_i(v_i;v_i,b_{-i};\rho)\ge \mathcal{U}_i(v_i;b_i,b_{-i};\rho)$.

In the \emph{stochastic} \problem, an adversary specifies a vector
    $\mu = (\mu_1 \LDOTS \mu_k)$
of CTRs (concealed from \A), then for each agent $i$ and round $t$, \realization $\rho_i(t)$ is chosen independently with mean $\mu_i$. Thus, an instance of the problem includes $\mu$ rather than a fixed \realization. For a given problem instance $\mathcal{I}$, let
    $i^* \in \argmax_i \mu_i\, v_i$,
then \emph{regret} on this instance is defined as
\begin{align}
R^\mathcal{I}(T)
    =    T\, v_{i^*} \mu_{i^*}  -
    \E\left[ \,\textstyle{\sum_{t=1}^T \sum_{i=1}^k}\, \mu_i\,v_i\;
        \A_i(b;\,\rho;\,t)\, \right].
\end{align}
For a given parameter \Vmax, the \emph{worst-case regret}\footnote{By abuse of notation, when clear from the context, the ``worst-case regret" is sometimes simply called ``regret".} $R(T; \Vmax)$ denotes the supremum of $R^\mathcal{I}(T)$ over all problem instances $\mathcal{I}$ in which all private values are at most $\Vmax$. Similarly, we define $R_{\delta}(T; \Vmax)$,
the \emph{worst-case $\delta$-regret}, by taking the supremum only on instances with $\delta$-gap.
%as defined next.
%If we reorder the agents such that agent $i$ has the $i$-th largest product $v_i\mu_i$, then an instance has $\delta$-gap if $\delta = (v_1\mu_1 - v_2\mu_2 )/  v_1 \mu_1$.

\OMIT{ %%%%%
An instance has $\delta$-gap if $v_{i(1)}\mu_{i(1)} - v_{i(2)}\mu_{i(2)}\ge \delta v_{i(1)}\mu_{i(1)}$ for $i(1),i(2)$ satisfying
$v_{i(1)}\mu_{i(1)}\geq v_{i(2)}\mu_{i(2)}\ge v_{j}\mu_{j}$ for any $j\neq i(1)$.
} %%%%

Most of our results are stated for \emph{non-degenerate} allocation rules, defined as follows. An interval is called \emph{non-degenerate} if it has positive length. Fix bid profile $b$, \realization $\rho$,  and rounds $t$ and $t'$ with $t\leq t'$. Let $i = \A(b;\rho;t)$ and $\rho'$ be the allocation obtained from $\rho$ by flipping the bit $\rho_i(t)$.  An allocation rule \A\ is \emph{non-degenerate} w.r.t. $(b,\rho,t,t')$ if there exists a non-degenerate interval $I$ containing $b_i$ such that
$$\A_i(x,b_{-i};\varphi;s) = \A_i(b;\varphi;s)\quad
\text{for each $\varphi\in\{\rho, \rho'\}$, each $s\in \{t,t'\}$, and all $x\in I$}.
$$
An allocation rule is \emph{non-degenerate} if it is non-degenerate w.r.t. each tuple $(b,\rho,t,t')$.

\OMIT{%%%%%%%%%%
Fixing bid profile $b$, \realization $\rho$,  agent $j$, and round $t$:
\begin{itemize}
\item An allocation rule is \emph{weakly non-degenerate} w.r.t. $(b,\rho,j,t)$ if there exists a non-degenerate interval $I$ containing $b_j$ such that
    $\A_j(x,b_{-j};\rho;t) = \A_j(b;\rho;t)$
for all $x\in I$.

% \item Predicate $P(b)$ means the following: for some round $t'$ we have 	$\A(b; \rho; t') \neq \A(b; \rho'; t')$, where \realization $\rho'$ is obtained from $\rho$ by flipping the bit $\rho_i(t)$.

\item An allocation rule for $k$ agents is \emph{non-degenerate} w.r.t. $(b,\rho,j,t)$
if it is weakly non-degenerate w.r.t. $(b,\rho,j,t)$ and moreover either $k=2$ or
 the following property holds:
\begin{itemize}
\item If $(b;\rho;j;i;t;t')$ is an influence-tuple for some $i$ and $t'$ then
$(x,b_{-j};\rho;j;i;t;t')$ is an influence-tuple for all $x$ in some non-degenerate interval containing $b_j$.
%\item If
%	$P(b_i, b_{-i})$
%then
% 	$P(x, b_{-i})$
%for all $x$ in some non-degenerate interval containing $b_i$.
\end{itemize}
\end{itemize}
An allocation rule is \emph{(weakly) non-degenerate} if it is (weakly) non-degenerate w.r.t. each tuple $(b,\rho,i,t)$.
} %%%%%%%%%%%%

 % preliminaries is a section again. (Alex: 11/23/08)
%\input{sec-2-scalefree} included in sec-k-truthful. Alex.
\section{Truthfulness characterization}
\label{sec:structural}

% We say that an MAB allocation rule \A\ is \emph{uniformly truthful} if there exists an MBA payment rule $\mathcal{P}$ such that the mechanism $(\A, \mathcal{P})$ is uniformly truthful.

Before presenting our characterization we begin by describing some related background.
The click allocation $\mathcal{C}$ is \emph{non-decreasing} if for each agent $i$, increasing her bid (and keeping everything else fixed) does not decrease $\mathcal{C}_i$. Prior work has established a characterization of truthful mechanisms for single-parameter domains (domains in which the private information of each agent is one-dimensional),
relating click allocation monotonicity and truthfulness (see below).
For our problem, this result is a characterization of MAB algorithms
that are truthful for a given \realization $\rho$, assuming that the \emph{entire} \realization $\rho$ can be used to compute payments (when computing payments one can use click information for every round and every agent, even if the agent was not shown at that round.) One of our main contributions is a characterization of MAB allocation rules that can be truthfully implemented when payment computation is restricted to only use clicks information of the actual impressions assigned by the allocation rule.

%Myerson~\cite{Myerson} and Archer and Tardos~\cite{ArcherTardos} proved the following characterization for truthful mechanisms:

\subsection{Monotonicity}

An MAB allocation rule \A\ is \emph{truthful with unrestricted payment computation} if it is truthful with a payment rule that can use the \emph{entire} \realization $\rho$ in it computation.
We next present the prior result characterizing truthful mechanisms with unrestricted payment computation.
\begin{theorem}[Myerson~\cite{Myerson}, Archer and Tardos~\cite{ArcherTardos}]
\OMIT{ %%% removed the footnote as per EC-camera-ready
\footnote{Archer and Tardos~\cite{ArcherTardos} was the first paper in the Theoretical Computer Science literature that presented a characterization of truthful mechanisms for single-parameter domains, in the context of machine scheduling.}
}
\label{thm:Myerson-characterization}
Let $(\A,\mathcal{P})$ be a normalized mechanism for the \problem. It is truthful with unrestricted payment computation if and only if for any given \realization $\rho$ the corresponding click-allocation $\mathcal{C}$ is non-decreasing and the payment rule is given by
\begin{align}\label{eq:Myerson-characterization}
\price_i(b_{i},b_{-i}; \rho) = b_i \cdot \mathcal{C}_i(b_{i},b_{-i}; \rho) -
        \textstyle{\int_0^{b_i}} \mathcal{C}_i(x,b_{-i}; \rho) \,dx. % + f_i(b_{-i})% ,\rho)
\end{align}
% where $f_i$ is arbitrary function.
% where $f_i(\cdot,\rho)$ is arbitrary function of $b_{-i}$ and the observed clicks in $\rho$.
% Moreover, the mechanism is truthful for $\rho$ \emph{and} satisfies voluntary participation if and only if~\refeq{eq:Myerson-characterization} holds with $f(\cdot)=0$.
% Moshe-10.27:
% This is NOT an if and only if with "voluntary participation" as we can pay each 10 before we start. this is iff with "normalized" payments.
\end{theorem}

We can now move to characterize truthful MAB mechanisms when the payment computation is restricted.
The following notation will be useful: for a given \realization $\rho$, let
    $\rho\xor\indicator(i,t)$,
be the \realization that coincides with $\rho$ everywhere, except that the bit $\rho_i(t)$ is flipped.

The first notable property of truthful mechanisms is a stronger version of monotonicity.
% MB5.2: I am repeating the definitions of "pointwise monotone" and "exploration-separated" from the intro "formally" upon the reviewer request
%AS2: OK, let's restate the definition. Just a little rewording.
Recall  (see Definition~\ref{def:pwm}) that an allocation rule \A\ is \emph{pointwise monotone} if for each \realization $\rho$, bid profile $b$, round $t$ and agent $i$, if $\A_i(b_i, b_{-i}; \rho; t)=1$ then $\A_i(b^+_i, b_{-i}; \rho; t)=1$ for any $b^+_i>b_i$. In words, increasing a bid cannot cause a loss of an impression.

\begin{lemma} \label{lemma:truthful-implies-pointwise-monotone}
Consider the \problem. Let $(\A,\mathcal{P})$ be a normalized truthful mechanism such that $\A$ is a non-degenerate deterministic allocation rule. Then $\A$ is pointwise-monotone.
\end{lemma}

\begin{proof}
For a contradiction, assume not. Then there is a \realization $\rho$, a bid profile $b$, a round $t$ and agent $i$ such that agent $i$ loses an impression in round $t$ by increasing her bid from $b_i$ to some larger value $b_i^+$. In other words, we have
    $ \A_i(b_i^+, b_{-i}; \rho; t) < \A_i(b_i, b_{-i}; \rho; t)$.
Without loss of generality, let us assume that there are no clicks after round $t$, that is $\rho_j(t')=0$ for any agent $j$ and any round $t'>t$ (since changes in $\rho$ after round $t$ does not affect anything before round $t$).

Let
    $\rho' =\rho\xor\indicator(i,t)$.
%be a \realization that coincides with $\rho$ everywhere, except the bit $\rho_i(t)$ is flipped.
The allocation in round $t$ cannot depend on this bit, so it must be the same for both \realizations. Now, for each \realization
    $\varphi \in \{\rho, \rho'\}$
the mechanism must be able to compute the price for agent $i$ when bids are $(b_i^{+},b_{-i})$. That involves computing the integral
     $I_i(\varphi) = \int_{x \leq b^+_i} \mathcal{C}_i(x, b_{-i};\varphi) \d x$
from~\refeq{eq:Myerson-characterization}. We claim that
    $I_i(\rho) \neq I_i(\rho')$.
However, the mechanism cannot distinguish between $\rho$ and $\rho'$ since they only differ in bit $(i,t)$ and agent $i$ does not get an impression in round $t$. This is a contradiction.

It remains to prove the claim. Without loss of generality, assume that $\rho_{i}(t)=0$ (otherwise interchange the role of $\rho$ and $\rho'$).  We first note that
    $\mathcal{C}_i(x, b_{-i}; \rho) \leq \mathcal{C}_i(x, b_{-i}; \rho')$
for every $x$. This is because everything is same in $\rho$ and $\rho'$ until round $t$ (so the impressions are same too), there are no clicks after round $t$, and in round $t$ the behavior of \A\ on the two \realizations can be different only if that agent $i$ gets an impression, in which case she is clicked under $\rho'$ and not clicked under $\rho$.

Since \A\ is non-degenerate, there exists a non-degenerate interval $I$ containing $b_i$ such that changing bid of agent $i$ to any value in this interval does not change the allocation at round $t$ (both for $\rho$ and for $\rho'$). For any $x\in I$  we have
    $\mathcal{C}_i(x, b_{-i}; \rho) < \mathcal{C}_i(x, b_{-i}; \rho')$,
where the difference is due to the click in round $t$. It follows that
    $I_i(\rho) < I_i(\rho')$. Claim proved.
Hence, the mechanism cannot be implemented truthfully.
\end{proof}

\subsection{Structural definitions}

Let us restate the structural definitions from the Introduction in a more detailed fashion.

\OMIT{ %%%%%%%
Recall from Definition~\ref{def:exploration-separated} that round $t$ is \emph{influential} for a given \realization $\rho$ if for some bid profile $b$ there exists a round $t'>t$ such that
    $\A(b;\rho;t') \neq \A(b;\rho\xor\indicator(j,t);t')$
for $j = \A(b;\rho;t)$. In words: changing the relevant part of the \realization at round $t$ affects the allocation in some future round $t'$. An allocation rule \A\ is called \emph{exploration-separated} if for any given \realization $\rho$ and round $t$ that is influential for $\rho$, it holds that $\A(b;\rho;t) = \A(b';\rho;t)$ for any two bid vectors $b,b'$ (allocation at $t$ does not depend on the bids).
} %%%%%%%%%

\begin{definition}\label{def:structure}
Fix \realization $\rho$, bid vector $b$, and round $t$.
\begin{itemize}

\item[(a)] Round $t$ is called {\em $(b;\rho)$-secured} from agent $i$ if $\A(b^+_i, b_{-i};\rho;t) = \A(b_i, b_{-i};\rho;t)$ for any $b^+_i>b_i$.

\item[(b)] Round $t$ is called {\em bid-independent} w.r.t.\ $\rho$ if the allocation $\A(b;\rho;t)$ is a constant function of $b$.

\item[(c)] Round $t$ is called {\em $(b;\rho)$-influential} if
 for some round
 $t'>t$ it holds that
    $\A(b;\rho;t')\neq \A(b;\rho';t')$
for \realization $\rho'= \rho\oplus \indicator(j,t)$ such that $j=\A(b;\rho;t)$.~\footnote{Note that \realizations $\rho$ and $\rho'$ are interchangeable.} In words: changing the relevant part of the \realization at round $t$ affects the allocation in some future round $t'$.

\item[(d)] In part (c), round $t'$ is called the {\em influenced round} and $j$ is called the {\em influencing agent} of round $t$.
The agent $i$ is called an {\em influenced agent} of round $t$ if
    $i\in \{ \A(b;\rho;t'),\, \A(b;\rho';t') \}$.

\item[(e)] Round $t$ is called \emph{influential} \wrt\ \realization $\rho$ if and only if it is $(b,\rho)$-influential for some $b$.
\end{itemize}
\end{definition}

\begin{definition}
Let $\A$ be a deterministic MAB allocation rule.
\begin{itemize}
\item $\A$ is called \emph{exploration-separated} if for every \realization $\rho$ and round $t$ that is influential for $\rho$, it holds that $\A(b;\rho;t) = \A(b';\rho;t)$ for any two bid vectors $b,b'$ (in words: allocation at round $t$ does not depend on the bids).

\item $\A$ is called \emph{weakly separated} if for every \realization $\rho$ and bid vector $b$, it holds that if round $t$ is $(b;\rho)$-influential with influenced agent $i$ then it is $(b;\rho)$-secured from $i$.
\end{itemize}
\end{definition}

\OMIT{The central property in our characterization is that each $(b,\rho)$-influential round is $(b,\rho)$-secured.}

\begin{observation}\label{obs:weakly-separated}
Any deterministic, exploration-separated MAB allocation rule is weakly separated.
\end{observation}

\begin{proof}
It follows from the definitions. Fix \realization $\rho$ and bid vector $b$, let $t$ be a $(b;\rho)$-influential round with influenced agent $i$. We need to show that $t$ is $(b;\rho)$-secured from $i$. Round $t$ is $(b;\rho)$-influential, thus influential w.r.t. $\rho$, thus (since  the allocation is exploration-separated) it is bid-independent w.r.t. $\rho$, thus agent $i$ cannot change allocation in round $t$ by increasing her bid.
\end{proof}

\begin{observation}\label{obs:exploration-separated}
Let $\A$ be a scale-free, weakly separated MAB allocation rule for two agents. Then $\A$ is exploration-separated.
\end{observation}

The proof of this observation is fairly straightforward, but it requires to carefully unwind the definitions. To provide some intuition with these definitions, we write it out in detail. 

\begin{proof}[Proof of Observation~\ref{obs:exploration-separated}]
Fix a \realization $\rho$ and round $t$ that is influential for $\rho$. Let $b,b'$ be two bid vectors. We need to conclude that 
    $\A(b;\rho;t) = \A(b';\rho;t)$.
    
By definition of ``influential round", there exists some bid vector $b^*$ such that $t$ is $(b^*,\rho)$-influential with influenced agent $i$. Since there are only two agents, the other agent is influenced, too. By definition of ``weakly separated", round $t$ is $(b^*,\rho)$-secured from both agents. By definition of ``secured", we have:
\begin{align}
\A(b^*;\rho;t) 
    &= \A(b^+_1, b^*_2;\, \rho;t) 
        \;\text{for any}\; b^+_1>b^*_1
        \label{eq:pf:obs:exploration-separated-1} \\
    &= \A(b^*_1, b^+_2;\, \rho;t)
        \;\text{for any}\; b^+_2>b^*_2.
        \label{eq:pf:obs:exploration-separated-2}
\end{align}

Let us prove that 
    $\A(b;\rho;t) = \A(b^*;\rho;t)$. 
We consider two cases. 
\begin{itemize}
\item Suppose 
    $b_1/b_2 \geq b^*_1/b^*_2$. 
Then by definition of ``scale-free", letting $\lambda = b^*_2/b_2$ we have
    $\A(b;\rho;t) = \A(\lambda b_1, b^*_2;\,\rho;t)$.
Since $\lambda b_1 > b^*_1$, then we are done by taking $b^+_1 = \lambda b_1$ and using \eqref{eq:pf:obs:exploration-separated-1}.

\item Suppose 
    $b_1/b_2 < b^*_1/b^*_2$.
Then by definition of ``scale-free", letting $\lambda = b^*_1/b_1$ we have 
    $\A(b;\rho;t) = \A(b^*_1, \lambda b_2;\,\rho;t)$.
Since $\lambda b_2 > b^*_2$, then we are done by taking $b^+_2 = \lambda b_2$ and using \eqref{eq:pf:obs:exploration-separated-2}.
\end{itemize}
Claim proved. Similarly, $\A(b';\rho;t) = \A(b^*;\rho;t)$.
\end{proof}

%%%%

\subsection{The two agents case (Theorem~\ref{thm:main-characterization-2-agents})}
%\subsection{The main structural implication for two agents}

The two-agent structural characterization in Theorem~\ref{thm:main-characterization-2-agents} follows from the general characterization in Theorem~\ref{thm:main-characterization}. More precisely, the ``if'' direction of Theorem~\ref{thm:main-characterization-2-agents} follows from the ``if'' direction of Theorem~\ref{thm:main-characterization} and Observation ~\ref{obs:weakly-separated}; the ``only if'' direction of Theorem~\ref{thm:main-characterization-2-agents} follows from the ``only if'' direction of Theorem~\ref{thm:main-characterization}  and Observation~\ref{obs:exploration-separated}.

The main structural implication in both theorems is that truthfulness implies the corresponding structural condition (either that the allocation rule is exploration separated or that it is weakly separated.) To illustrate the ideas behind this implication, we prove the two-agent case directly.

\begin{proposition}
\label{lemma:truthful-implies-exploration-separated-2-players-scalefree}
Consider the \problem\ with two agents. Let $\A$ be a non-degenerate scale-free deterministic allocation rule.
If $(\A,\mathcal{P})$ is a normalized truthful mechanism for some $\mathcal{P}$, then it is exploration separated.
%Consider the \problem\ with two agents. Let $(\A,\mathcal{P})$ be a normalized truthful mechanism such that $\A$ is a non-degenerate deterministic allocation rule. If $\A$ is scale-free then it is exploration-separated.
\end{proposition}

\begin{proof}
Assume $\A$ is
%scale-free but
not exploration-separated. Then there is a \emph{counterexample} $(\rho,t)$: a \realization $\rho$ and a round $t$ such that round $t$ is influential and allocation in round $t$ depends on bids. We want to prove that this leads to a contradiction.

Let us pick a counterexample $(\rho,t)$ with some useful properties. Since round $t$ is influential, there exists a \realization $\rho$ and bid profile $b$ such that the allocation at some round $t'>t$ (the \emph{influenced} round) is different under \realization $\rho$ and another \realization
    $\rho'=\rho\xor\indicator(j,t)$,
where
    $j = \A(b;\rho;t)$
is the agent chosen at round $t$ under $\rho$. Without loss of generality, let us pick a counterexample with minimum value of $t'$ over all choices of $(b,\rho,t)$. For ease of exposition, from this point on let us assume that $j=2$. For the counterexample we can also assume that $\rho_1(t')=1$, and that there are no clicks after round $t'$, that is
    $\rho_l(t'') = \rho'_{l}(t'')=0$ for all $t''>t'$ and for all $l\in\{1,2\}$.

We know that the allocation in round $t$ depends on bids. This means that agent $1$ gets an impression in round $t$ for some bid profile $\hat{b}=(\hat{b}_1, \hat{b}_2)$ under \realization $\rho$, that is
    $\A(\hat{b}; \rho; t)=1$.
As the mechanism is scale-free this means that, denoting
    $b^+_1 = \hat{b}_1\, b_2/\hat{b}_2$
we have
    $\A(b^+_1, b_2; \rho; t)=1$.
Since $\A(b_1, b_2; \rho; t)=2$ and $\A(b^+_1, b_2; \rho; t)=1$,
pointwise monotonicity (Lemma~\ref{lemma:truthful-implies-pointwise-monotone}) implies that $b^+_1 >b_1$. We conclude that there exists a bid $b_1^+ > b_1$ for agent $1$ such that
    $\A(b_1^+, b_2; \rho; t)=1$.

\OMIT{%%%%
In the pricing equation~\refeq{eq:Myerson-characterization} we have
    $\mathcal{C}_1(b_1^+, b_2; \rho)=\mathcal{C}_1(b_1^+, b_2; \rho')$,
since the differing bit $\rho_2(t)$ is not observed when bid of agent $1$ is $b_1^+$.
} %%%%%%%

Now, the mechanism needs to compute prices for agent $1$ for bids $(b_1^+,\, b_2)$ under \realizations $\rho$ and $\rho'$, that is
    $\mathcal{P}_1(b_1^{+}, b_2; \rho)$ and $\mathcal{P}_1(b_{i}^{+}, b_2; \rho')$.
Therefore, the mechanism needs to compute the integral
     $I_1(\varphi) = \int_{x \leq b^+_1} \mathcal{C}_1(x, b_2;\varphi) \d x$
for both \realizations $\varphi \in \{\rho, \rho'\}$.

First of all, for all $x\leq b_1^+$ and for all $t''<t'$,
    $\A(x,b_2; \rho; t'')= \A(x,b_2;\rho'; t'')$,
since otherwise the minimality of $t'$ will be violated. The only difference in the allocation can occur in round $t'$.

Let us assume
    $\A_1(b_1,b_2;\rho;t')<\A_1(b_1,b_2;\rho',t')$
(otherwise, we can swap $\rho$ and $\rho'$). We make the claim that for all bids $x\leq b_1^+$ of agent $1$, the influence of round $t$ on round $t'$ is in the same ``direction'':
\begin{align}\label{eq:influences-in-same-direction}
\A_1(x,b_2;\rho;t')\leq \A_1(x,b_2; \rho'; t')
    \text{~~for all~~} x\leq b_1^+.
\end{align}
Suppose~\refeq{eq:influences-in-same-direction} does not hold. Then there is an $x< b_1^+$ such that
    $1=\A_1(x,b_2;\rho;t')> \A_1(x,b_2;\rho';t')=0$.
(Note that we have used the fact that the mechanism is deterministic.) If $x<b_1$ then pointwise monotonicity is violated under \realization $\rho$, since
    $\A_1(x,b_2;\rho;t')>A_1(b_{1},b_2;\rho;t')$;
otherwise it is violated under \realization $\rho'$, giving a contradiction in both cases. The claim~\refeq{eq:influences-in-same-direction} follows.

Since \A\ is non-degenerate, there exists a non-degenerate interval $I$ containing $b_i$ such that if agent $1$ bids any value $x\in I$ then
  $\A_1(x,b_2;\rho;t')< \A_1(x,b_2;\rho';t')$.
Now by~\refeq{eq:influences-in-same-direction} it follows that
    $I_1(\rho) < I_2(\rho')$.
However, the mechanism cannot distinguish between $\rho$ and $\rho'$ when the bid of agent $1$ is $b_1^+$, since the differing bit $\rho_2(t)$ is not observed. Therefore the mechanism cannot compute prices, contradiction.
\end{proof}

%%%%%%%%%%%%%%%%%%%%%%
\subsection{The general case (Theorem~\ref{thm:main-characterization})}

Let us prove the general characterization (Theorem~\ref{thm:main-characterization}). We restate it here for convenience.

\begin{theorem*}[Theorem~\ref{thm:main-characterization}, restated]
%\label{thm:characterization-k-non-scalefree}
Consider the \problem. Let \A\ be a non-degenerate deterministic allocation rule. Then a mechanism $(\A,\mathcal{P})$ is normalized and truthful for some payment rule $\mathcal{P}$ if and only if $\A$ is pointwise monotone and weakly separated.
\end{theorem*}

\begin{proof}[Proof of Theorem~\ref{thm:main-characterization}: the ``only if'' direction]
Suppose $(\A,\mathcal{P})$ be a normalized truthful mechanism, for some payment rule $\mathcal{P}$. Then $\A$ is pointwise-monotone by Lemma~\ref{lemma:truthful-implies-pointwise-monotone}. The fact that \A\ is weakly separated is proved similarly to Proposition~\ref{lemma:truthful-implies-exploration-separated-2-players-scalefree}, albeit with a few extra details.

Assume $\A$ is not weakly separated. Then there is a \emph{counterexample} $(\rho,b,t,t',i)$: a \realization $\rho$, bid vector $b$, rounds $t,t'$ and agent $i$ such that round $t$ is $(b;\rho)$-influential with influenced agent $i$ and influenced round $t'$  and it does not holds that round $t$ is $(b;\rho)$-secured from $i$. We prove that this leads to a contradiction..

Let us pick a counterexample $(\rho,b,t,t',i)$ with a minimum value of $t'$ over all choices of $(\rho,b,t,i)$. Without loss of generality, let us assume that $\rho_i(t')=1$ and $\rho_j(t'')=0$ for all $t''>t'$ and for all agents $j$.

Let $j = \A(b;\rho;t)$. As it does not holds that round $t$ is $(b;\rho)$-secured from $i$, this means that $j\neq i$, and
there exists a bid $b^+_i>b_i$ such that $\A(b^+_i,b_{-i};\rho;t)\neq j$.
% Thus conclude that there exists a bid $b_i^+ > b_i$ for agent $i\not=\A(b;\rho;t)$, such that $\A(b_i^+, b_{-i}; \rho; t)\neq j$.

Let $\rho'=\rho\xor\indicator(j,t)$. The mechanism needs to compute prices for agent $i$ when her bid is $b_i^+$ under \realizations $\rho$ and $\rho'$, that is to compute
    $\price_i(b_i^{+}, b_{-i}; \rho)$
and
    $\price_i(b_{i}^{+}, b_{-i}; \rho')$.
Therefore, the mechanism needs to compute the integral
     $I_i(\varphi) = \int_{x \leq b^+_1} \mathcal{C}_i(x, b_{-i};\varphi) \d x$
for both \realizations $\varphi \in \{\rho, \rho'\}$.

First of all, for all $x\leq b_i^+$ and for all $t''<t'$, $\A_i(x,b_{-i}; \rho; t'')=\A_i(x,b_{-i};\rho'; t'')$. If not,then the minimality of $t'$ will be violated. This is because, if there were such an $x$ and $t''<t'$ with $\A_i(x,b_{-i};\rho;t'')\not= \A_i(x,b_{-i};\rho';t'')$, then round $t$ will still be $(b,\rho)$-influential with influenced agent $i$, and influenced round $t''<t'$, violating the minimality of $t''$. Therefore, when we decrease the bid of agent $i$, the only difference in the allocation can occur at time round $t'$.

As $i$ is the influenced agent at round $t'$ it must hold that $\A_i(b_i,b_{-i};\rho;t')\neq \A_i(b_i,b_{-i};\rho',t')$.
Let us assume $0=\A_i(b_i,b_{-i};\rho;t') < \A_i(b_i,b_{-i};\rho',t')=1$ (otherwise, we can swap $\rho$ and $\rho'$). Note that we have made use of the fact that the mechanism is deterministic. Let us make the the claim that for all bids $x\leq b^+_i$ the influence of round $t$ on round $t'$ is in the same ``direction.''
\begin{align}\label{eq:influences-in-same-direction-k-players}
\A_i(x,b_{-i};\rho;t')\leq \A_i(x,b_{-i}; \rho'; t')
    \text{~~for all $x\leq b_i^+$}.
\end{align}
Suppose~\refeq{eq:influences-in-same-direction-k-players} does not hold. Then there is an $x\leq b_i^+$ such that $1=\A_i(x,b_{-i};\rho;t')>\A_i(x,b_{-i};\rho';t')=0$. (Note that we have used the fact that the mechanism is deterministic.) If $x>b_i$, then pointwise monotonicity is violated in $\rho'$, since $0=\A_i(x,b_{-i};\rho';t')<\A_i(b_{i},b_{-i};\rho';t')=1$. If $x<b_i$ on the other hand, then the pointwise-monotonicity is violated in $\rho$, since $1=\A_i(x,b_{-i};\rho;t')>\A_i(b_i,b_{-i};\rho;t')=0$, giving a contradiction in both cases. The claim~\refeq{eq:influences-in-same-direction-k-players} follows.

By the non-degeneracy of \A, there exists a non-degenerate interval $I$ containing $b_i$ such that
\begin{align}\label{eq:influence-positive}
\A_i(x,b_{-i};\rho;t')<\A_i(x,b_{-i};\rho';t')
\text{~~for all $x\in I$}.
\end{align}
By~\refeq{eq:influences-in-same-direction-k-players} and ~\refeq{eq:influence-positive} it follows that
    $I_i(\rho) < I_i(\rho') $.
However, the mechanism cannot distinguish between $\rho$ and $\rho'$ when agent $i$'s bid is $b_i^+$, since the differing bit $\rho_j(t)$ is not seen. Contradiction.
\end{proof}

\begin{proof}[Proof of Theorem~\ref{thm:main-characterization}: the ``if'' direction]
Let $\A$ be a deterministic allocation rule which is pointwise monotone and weakly separated. We need to provide a payment rule $\mathcal{P}$ such that the resulting mechanism $(\A,\mathcal{P})$ is truthful and normalized. Since \A\ is pointwise monotone, it immediately follows that it is monotone (i.e., as an agent increases her bid, the number of clicks that she gets cannot decrease). Therefore it follows from Theorem~\ref{thm:Myerson-characterization} that mechanism $(\A,\mathcal{P})$ is truthful and normalized if and only if $\mathcal{P}$ is given by~\refeq{eq:Myerson-characterization}. We need to show that $\mathcal{P}$ can be computed using only the knowledge of the clicks (bits from the \realization) that were revealed during the execution of \A.

Assume we want to compute the payment for agent $i$ in bid profile $(b_i, b_{-i})$ and \realization $\rho$. We will prove that we can compute $\mathcal{C}_i(x):=\mathcal{C}_i(x,b_{-i};\rho)$ for all $x\leq b_i$.
To compute $\mathcal{C}_i(x)$, we show that it is possible to simulate the execution of the mechanism with $\bid_i=x$.
In some rounds, the agent $i$ loses an impression, and in others it retains the impression (pointwise monotonicity ensures that agent $i$ cannot gain an impression when decreasing her bid). In rounds that it loses an impression, the mechanism does not observe the bits of $\rho$ in those rounds, so we prove that those bits are {\em irrelevant} while computing $\mathcal{C}_i(x)$. In other words, while running with $\bid_i=x$, if mechanism needs to observe the bit that was not revealed when running with $\bid_i=b_i$, we arbitrarily put that bit equal to $1$ and simulate the execution of \A. We want to prove that this computes $\mathcal{C}_i(x)$ correctly.

%    (that is, $A(b;\rho;t_{j})= i \neq A(x,b_{-i};\rho;t_{j})$ for any $j\in \{1,2,\ldots,n\}$.)

Let $t_1 < t_2 < \dots < t_n$ be the rounds in which agent $i$ did not get an impression while bidding $x$, but did get an impression while bidding $b_i$. Let $\rho^0:=\rho$, and let us define \realization $\rho^l$ inductively for every $l \in [n]$ by setting
    $\rho^l:=\rho^{l-1} \xor \indicator(j_l,\,t_l)$,
where
    $j_l=\A(x,b_{-i};\rho^{l-1};t_l)$
is the agent that got the impression at round $t_l$ with \realization $\rho^{l-1}$ and bids $(x,b_{-i})$.

First, we claim that $j_l \neq i$ for any $l$. Indeed, suppose not, and pick the smallest $l$ such that $j_{l+1} = i$. Then $t_l$ is a $(x, b_{-i};\, \rho^l)$-influential round, with influenced agent $j_{l+1} = i$. Thus $t_l$ is $(x, b_{-i};\, \rho^l)$-secured from $i$. Since
    $\A(x,b_{-i};\, \rho^l; t_l) = \A(x,b_{-i};\, \rho^{l-1}; t_l) = j_l \neq i$
by minimality of $l$, agent $i$ does not get an impression in round $t_l$ if she raises her bid to $b_i$. That is,
    $\A(b; \rho^l; t_l) \neq i$.
However, the changes in \realizations $\rho^0 \LDOTS \rho^{l-1}$ only concern the rounds in which agent $i$ is chosen, so they are not seen by the allocation if the bid profile is $b$ (to prove this formally, use induction). Thus,
    $\A(b; \rho^l; t_l) = \A(b;\rho; t_l) = i $,
contradiction. Claim proved. It follows that $\A(b;\rho; t_l) = i $ for each $l$. (This is because by induction, the change from $\rho^{l-1}$ to $\rho^l$ is not seen by the allocation if the bid profile is $b$.)

We claim that
    $\A_i(x,b_{-i};\rho;t')=\A_i(x,b_{-i};\rho^n;t')$
for every round $t'$, which will prove the theorem. If not, then there exists $l$ such that
    $\A_i(x,b_{-i};\rho^l;t')\neq \A_i(x,b_{-i};\rho^{l-1};t')$
for some $t'$ (and of course $t'>t_l$).
%Consider the minimal such $l$.
Round $t_l$ is thus $(x,b_{-i};\rho^l)$-influential with influenced round $t'$ and influenced agent $i$.
Moreover, the influencing agent of that round is $j_l$, and we already proved that $j_l \neq i$.  Since round $t_l$ is $(x,b_{-i};\rho^l)$-secured from agent $i$ due to the ``weakly separated'' condition, it follows that agent $i$ does not get an impression in round $t_l$ if she raises her bid to $b_i$. That is,
    $\A(b; \rho^l; t_l) \neq i$, contradiction.
\end{proof}

\OMIT{ %%%%%%%%%%%
That is, $A_i(x,b_{-i};\rho^{l+1};t_{l+1})=A_i(x,b_{-i};\rho^{l};t_{l+1})\not=i$. This follows from the following fact, which can be proved by an induction on $m$: for all $m\leq l$, $A_i(x,b_{-i};\rho^{m};t_{l+1})\not=i$. This is true for $m=0$ (assumption at the beginning about rounds $t_{l+1}$). If $A_i(x,b_{-i};\rho^m;t_{l+1})=0$ and $A_i(x,b_{-i};\rho^{m+1};t_{l+1})=1$ for some $m\leq l$, then round $t_{m+1}$ is $(x,b_{-i},\rho^{m})$-influential round with influenced agent $i$ and influenced round $t_{l+1}$ and round $t_{m+1}$ is not $(x,b_{-i};\rho^{m+1})$-secured from $i$ ($i$ can increase her bid to $b_i$ to get the impression).

influenced agent $i$. Moreover, we just proved that the influencing agent in round $t_{l+1}$ is not $i$.
To derive a contradiction we show that round $t_{l+1}$ is not $(x,b_{-i};\rho^{l+1})$-secured from agent $i$.
Recall that $A(b_i,b_{-i};\rho;t_{l+1})= i$. It also must be the case that $A(b_i,b_{-i};\rho^{l+1};t_{l+1})= i$. This is so as $i$ gets all impressions of rounds $t_{j}$ for any $j\in \{1,2,\ldots,l+1\}$, and there is no difference in the clicks of $i$ between $\rho$ and $\rho^{l+1}$ at rounds up to $t_{l+1}$.
Thus we have seen that $A(b_i,b_{-i};\rho^{l+1};t_{l+1})= i$ while $A(x,b_{-i};\rho^{l+1};t_{l+1})=j_{l+1}\neq i$, which implies that
round $t_{l+1}$ is not $(x,b_{-i};\rho^{l+1})$-secured from agent $i$.
This is a contradiction to the condition presented in the lemma statement.

Recall that $A(b_i,b_{-i};\rho;t_{l+1})= i$. It also must be the case that $A(b_i,b_{-i};\rho^{l+1};t_{l+1})= i$. This is so as $i$ gets all impressions of rounds $t_{j}$ for any $j\in \{1,2,\ldots,l+1\}$, and there is no difference in the clicks of $i$ between $\rho$ and $\rho^{l+1}$ at rounds up to $t_{l+1}$.
Thus we have seen that $A(b_i,b_{-i};\rho^{l+1};t_{l+1})= i$ while $A(x,b_{-i};\rho^{l+1};t_{l+1})=j_{l+1}\neq i$, which implies that
round $t_{l+1}$ is not $(x,b_{-i};\rho^{l+1})$-secured from agent $i$.
This is a contradiction to the condition presented in the lemma statement.
\

This is so as $l$ is the minimal one that satisfies the condition, thus it must be that $t_{l+1}\neq i$ (as $A(x,b_{-i};\rho;t_{j})\neq i$ for any $j\in \{1,2,\ldots,n\}$ and we look at the minimal $l$ such that there is a change in the allocation to $i$).
} %%%%%%%%%%%%

%\begin{lemma}
%\label{lem:scallfree-and-iia-implies-condition}
%Consider the \problem\ with $k$ agents. Let \A\ be an MAB (deterministic) allocation rule which is non-degenerate.
% Assume that \A\ is scalefree and satisfies IIA.
% Then, for any \realization $\rho$ and bids $b$, if round $t$ is $(b;\rho)$-influential with influenced agent $i$ then it is  $(b;\rho)$-secured from $i$.
% \end{lemma}

\OMIT{ %%% re-worded the para as per EC-camera-ready
Note that we have proven the main characterization result (Theorem~\ref{thm:main-characterization}) for the case of two agents, because for two agents, it is not hard to see that a scale-free allocation is exploration-separated if and only if it is weakly separated, and also IIA trivially holds for two agents.
}

Let us argue that the non-degeneracy assumption in Theorem~\ref{thm:main-characterization} is indeed necessary.

\begin{claim}\label{cl:degeneracy}
There exists a deterministic mechanism $(\A,\mathcal{P})$ for two agents that is truthful and normalized, such that the allocation rule $\A$ is pointwise monotone, scale-free and yet \emph{not} weakly separated.
\end{claim}

\begin{proof}
There are only two rounds. Agent $1$ allocated at round 1 if and only if $b_1\geq b_2$. Agent $1$ allocated at round $2$ if $b_1 > b_2$ or if
    $b_1=b_2$ and $\rho_1(1)=1$;
otherwise agent 2 is shown. This completes the description of the allocation rule. To obtain a payment rule $\mathcal{P}$ which makes the mechanism normalized and truthful, consider an alternate allocation rule $\A'$ which in each round selects agent $1$ if and only if $b_1\geq b_2$. (Note that $\A' = \A$ except when $b_1 = b_2$.) Use Theorem~\ref{thm:main-characterization} for $\A'$ to obtain a normalized truthful mechanism $(\A', \mathcal{P}')$, and set $\mathcal{P} = \mathcal{P}'$. The payment rule $\mathcal{P}$ is well-defined since the observed clicks for $\mathcal{P}$ and $\mathcal{P}'$ coincide unless $b_1=b_2$, in which case both payment rules charge $0$ to both agents. The resulting mechanism $(\A, \mathcal{P})$ is normalized and truthful because the integral in~\refeq{eq:Myerson-characterization} remains the same even if we change the value at a single point. It is easy to see that the allocation rule $\A$ has all the claimed properties; it fails to be non-degenerate because round $t$ is influential only when $b_1=b_2$.
\end{proof}

%%%%%%%%%%%%%%%%%%%%
\subsection{Scalefree and IIA allocation rules}

We show that under the right assumptions, an MAB allocation rule is exploration-separated if and only if it is weakly separated.

\begin{lemma}
\label{lem:scalefree-and-iia-implies-equivalent-conditions}
Consider the \problem. Let \A\ be a non-degenerate deterministic allocation rule which is scalefree, pointwise monotone, and satisfies IIA. Then it is exploration-separated if and only if it is weakly separated.
\end{lemma}

The proof of Lemma~\ref{lem:scalefree-and-iia-implies-equivalent-conditions} is very technical. We precede it with a proof sketch. To preserve the flow, we place the full proof in Appendix~\ref{sec:IIA}.
% ,which can be skipped in the first reading.

\begin{figure}[t]
%\vspace{-7mm}
\begin{center}
\includegraphics[width=5in]{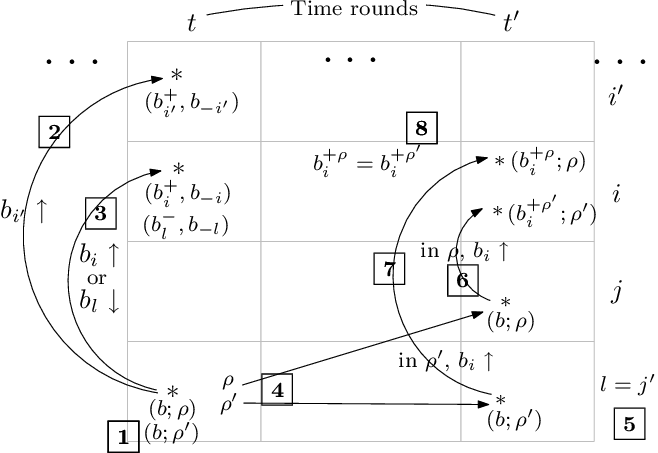}
\end{center}
\caption{\small This figure explains all the steps in the proof of Lemma~\ref{lem:scalefree-and-iia-implies-equivalent-conditions}. The rows correspond to agents (whose identity is shown on the right side), and columns correspond to time rounds. The asterisks show the impressions. The arrows show how the impressions get {\em transferred}, and labels on the arrows show what causes the transfer. In labels, ``in $\rho$, $b_i\uparrow$'' denotes that a particular transfer of impression is caused in \realization $\rho$ when bid $b_i$ in increased.}
\label{figure:proof-of-k-player-scale-free-iia}
\end{figure}

\begin{proof}[Proof Sketch.]
We sketch the proof of Lemma~\ref{lem:scalefree-and-iia-implies-equivalent-conditions} at a {\em very} high level.
The ``only if'' direction was observed in Observation~\ref{obs:weakly-separated}; we focus on the ``if'' direction. Let $\A$ be a weakly-separated mechanism. We prove by a contradiction that it is exploration-separated. If not, then there is a \realization $\rho$ and a round $t$ such that $t$	 is influencial \wrt $\rho$ as well as not bid-dependent \wrt $\rho$. Let round $t$ be influencial with bid vector $b$, influencing agent $l$, and influenced agents $j$ and $j'\not=j$ in influenced round $t'$ (see $\InFigureLAtRoundT$ in Figure~\ref{figure:proof-of-k-player-scale-free-iia}; all boxed numbers in this sketch will refer to this figure).

From the assumption, $t$ is not bid-dependent \wrt $\rho$, which means that there exists a bid profile $b'$ such that $i'\not=l$ is selected in round $t$ with bids $b'$. Using scalefreeness, IIA, and pointwise-monotonicity, we can prove that there exists a sufficiently large bid $b_{i'}^{+}$ of agent $i'$ such that she gets an impression in round $t$ with bids $(b_{i'}^{+},b_{-i'})$ (see \InFigureIPrimeAtRoundTForLargeBid). Using the properties of the mechanism, it can further be proved that there is an agent $i$ such that she gets the impression in round $t$ when either $i$ increases her bid, {\em or} $l$ decreases her bid (see $\InFigureIIncreasesHerBid$). When $i$ increases her bid to $b_i^{+}$, she also gets an impression in round $t'$, since impressions cannot differ in round $t'$ in the case when $l$ is not selected in round $t$ and they must get transferred from $j$ and $j'$ to {\em somebody} in round $t'$, and IIA implies that this {\em somebody} should be $i$.

Recall that two different agents $j$ and $j'$ get the impression in round $t'$ under $\rho$ and $\rho'$ respectively (see $\InFigureRoundTIsInfluential$).
We prove that either agent $j'$ or agent $j$ must be equal to $l$ (this is done by looking at how the allocation in round $t'$ changes when $l$ decreases her bid). Let us break the symmetry and assume $j'=l$ (see box $\InFigureLIsJOrJPrime$). It is also easy to see that when $i$ increases her bid, impression in round $t'$ get transferred to her in $\rho$ (at some minimum value $b_i^{+\rho}$, see $\InFigureBIPlusRho$), and impression in round $t'$ gets transferred to her also in $\rho'$ (as some possibly different minimum value $b_i^{+\rho'}$, see $\InFigureBIPlusRhoPrime$). Using the assumptions of weakly-separatedness, we prove that $b_i^{+\rho}=b_i^{+\rho'}$ (see $\InFigureBIPlusRhoEqualsBIPlusRhoPrime$). This can be proved by observing that $b_i^+\geq \max\{b_i^{+\rho},b_i^{+\rho'}\}$, and then using weakly-separatedness of $\A$. Since these two bids were at a ``threshold value'' (these were the minimum values of bids to have transferred the impression in $\rho$ and $\rho'$ from $j$ and $l$ respectively), we are able to prove that the ratio of $b_j/b_l$ must be some fixed number dependent on $\rho$, $\rho'$, and $t'$. In particular, it follows that $b_l$ belongs to a finite set $S(b_{-l})$ which depends only on $b_{-l}$. However, by non-degeneracy of \A\ there must be infinitely many such $b_l$'s, which leads to a contradiction.
\end{proof}

\OMIT{But this contradicts the assumptions of non-degeneracy. We can decrease the bid of $b_j$ by a small amount (say $\epsilon$, which exists by the non-degeneracy assumption and the fact that are finitely many rounds) so that $j$'s allocation does not change in any round, either in $\rho$ or in $\rho'$. Using IIA, it follows that {\em nobody}'s allocation changes in any round. Now we can carry out the same argument as above for bid vector $(b_{j}-\epsilon,b_{-j})$, and it will prove that the ratio $(b_j-\epsilon)/b_l=b_j/b_l$, a contradiction since $\epsilon\not=0$. }

\section{Lower bounds on regret}
\label{sec:regret}

In this section we use structural results from the previous section to derive lower bounds on regret.

\begin{theorem}\label{thm:LB-main}
Consider the stochastic \problem\ with $k$ agents. Let \A\ be an exploration-separated deterministic allocation rule. Then its regret is
   $R(T;\, \Vmax) = \Omega(\Vmax\, k^{1/3}\, T^{2/3})$.
\end{theorem}

%    R_\delta(T;\, \Vmax) &= \Omega(\tfrac{1}{\delta}\,\Vmax\, T^\gamma)

Let
    $\vec{\mu}_0 = (\tfrac12 \LDOTS \tfrac12) \in [0,1]^k$
be the vector of CTRs in which for each agent the CTR is $\tfrac12$. For each agent $i$, let $\vec{\mu}_i  = (\mu_{i1}, \dots, \mu_{ik})\in [0,1]^k$ be the vector of CTRs in which agent $i$ has CTR
    $\mu_{ii} = \tfrac12+\eps$, $\eps=k^{1/3}\, T^{-1/3} $,
and every other agent $j\neq i$ has CTR $\mu_{ij} = \tfrac12$. As a notational convention, denote by $\mathbb{P}_i[\cdot]$ and $\mathbb{E}_i[\cdot]$ respectively the probability and expectation induced by the algorithm when clicks are given by $\vec{\mu}_i$.
Let $\mathcal{I}_i$ be the problem instance in which CTRs are given by $\vec{\mu}_i$ and all bids are \Vmax. For each agent $i$, let $\mathcal{J}_i$ be the problem instance in which CTRs are given by $\vec{\mu}_0$, the bid of agent $i$ is \Vmax, and the bids of all other agents are $\Vmax/2$. We will show that for any exploration-separated deterministic allocation rule \A, one of these $2k$ instances causes high regret.

Let $N_i$ be the number of bid-independent rounds in which agent $i$ is selected. Note that $N_i$ does not depend on the bids. It is a random variable in the probability space induced by the clicks; its distribution is completely specified by the CTRs. We show that (in a certain sense) the allocation cannot distinguish between $\vec{\mu}_0$ and $\vec{\mu}_i$ if $N_i$ is too small. Specifically, let $\A_t$ be the allocation in round $t$. Once the bids are fixed, this is a random variable in the probability space induced by the clicks. For a given set $S$ of agents, we consider the event $\{\A_t\in S\}$ for some fixed round $t$, and upper-bound the difference between the probability of this event under $\vec{\mu}_0$ and $\vec{\mu}_i$ in terms of $\mathbb{E}_i[N_i]$, in the following crucial claim, which is proved in Section~\ref{app:regret} via relative entropy techniques.

\begin{claim}\label{cl:LB-crux}
For any fixed vector of bids, each round $t$, each agent $i$ and each set of agents $S$, we have
\begin{align}\label{eq:LB-crux}
  |\, \mathbb{P}_0[\A_t \in S] - \mathbb{P}_i[\A_t \in S] \,|
    \leq O(\eps^2\; \mathbb{E}_0[N_i]).
\end{align}
\end{claim}

%This crucial claim is proved in Appendix~\ref{app:regret} via relative entropy techniques.

\begin{proofof}{Theorem~\ref{thm:LB-main}}
Fix a positive constant $\beta$ to be specified later. Consider the case $k=2$ first. If
    $\mathbb{E}_0[N_i] > \beta\, T^{2/3}$
for some agent $i$, then on the problem instance $\mathcal{J}_i$, regret is $\Omega(T^{2/3})$. So without loss of generality let us assume
    $\mathbb{E}_0[N_i] \leq \beta\, T^{2/3}$
for each agent $i$. Then, plugging in the values for $\eps$ and $\mathbb{E}_0[N_i]$, the right-hand side of~\refeq{eq:LB-crux} is at most $O(\beta)$. Take $\beta$ so that the right-hand side of~\refeq{eq:LB-crux} is at most $\tfrac14$. For each round $t$ there is an agent $i$ such that
    $\mathbb{P}_0[\A_t \neq i] \geq \tfrac12$.
Then
    $\mathbb{P}_i[\A_t \neq i] \geq \tfrac14$
by Claim~\ref{cl:LB-crux}, and therefore in this round algorithm \A\ incurs regret $\Omega(\eps\,\Vmax)$ under problem instance $\mathcal{I}_i$. By Pigeonhole Principle there exists an $i$ such that this happens for at least half of the rounds $t$, which gives the desired lower-bound.

Case $k\geq 3$ requires a different (and somewhat more complicated) argument. Let
    $R = \beta\, k^{1/3}\, T^{2/3} $
and $N$ be the number of bid-independent rounds. Assume $\mathbb{E}_0[N] > R$. Then
    $\mathbb{E}_0[N_i] \leq \tfrac1k\,\mathbb{E}_0[N]$
for some agent $i$. For the problem instance $\mathcal{J}_i$ there are, in expectation,
    $E[N-N_i] = \Omega(R)$
bid-independent rounds in which agent $i$ is not selected; each of which contributes $\Omega(\Vmax)$ to regret, so the total regret is $\Omega(\Vmax\,R)$.

From now on assume that
    $\mathbb{E}_0[N] \leq R$.
Note that by Pigeonhole Principle, there are more than $\tfrac{k}{2}$ agents $i$ such that
    $\mathbb{E}_0[N_i] \leq 2R/k$.
Furthermore, let us say that an agent $i$ is \emph{good} if
    $\mathbb{P}_0[\A_t = i] \leq \tfrac45$
for more than $T/6$ different rounds $t$. We claim that there are more than $\tfrac{k}{2}$ good agents.  Suppose not. If agent $i$ is not good then
    $\mathbb{P}_0[\A_t = i] > \tfrac45$ for at least $\tfrac56 T$
different rounds $t$, so if there are  at least $k/2$ such agents then
\begin{align*}
T = \textstyle{\sum_{t=1}^T \sum_{i=1}^k } \mathbb{P}_0[\A_t = i]
    > \tfrac{k}{2}\times (\tfrac56 T)\times \tfrac45
    \geq kT/3 \geq T,
\end{align*}
contradiction. Claim proved. It follows that there exists a good agent $i$ such that
    $\mathbb{E}_0[N_i] \leq 2R/k$.
Therefore the right-hand side of~\refeq{eq:LB-crux} is at most $O(\beta)$. Pick $\beta$ so that the right-hand side of~\refeq{eq:LB-crux} is at most $\tfrac{1}{10}$. Then by Claim~\ref{cl:LB-crux} for at least $T/6$ different rounds $t$ we have
    $\mathbb{P}_i[\A_t = i] \leq \tfrac{9}{10}$.
In each such round, if agent $i$ is not selected then algorithm \A\ incurs regret $\Omega(\eps\,\Vmax)$ on problem instance $\mathcal{I}_i$. Therefore, the (total) regret of $\A$ on problem instance $\mathcal{I}_i$ is
    $\Omega(\eps\,\Vmax\,T) = \Omega(\Vmax\, k^{1/3}\, T^{2/3})$.
\end{proofof}

%%%%%%%%%%%%%%%%%%

\begin{theorem}\label{thm:LB-delta}
In the setting of Theorem~\ref{thm:LB-main}, fix $k$ and $\Vmax$ and assume that
    $R(T;\, \Vmax) =O(\Vmax\, T^\gamma)$
for some $\gamma<1$. Then for every fixed $\delta\leq \tfrac14$ and $\lambda < 2(1-\gamma)$
we have
    $R_\delta(T;\, \Vmax) = \Omega(\delta\,\Vmax\,T^\lambda)$.
\end{theorem}
\begin{proof}
Fix $\lambda \in (0,\, 2(1-\gamma))$.
Redefine $\vec{\mu}_i$'s with respect to a different $\eps$, namely $\eps = T^{-\lambda/2}$. Define the problem instances $\mathcal{I}_i$ in the same way as before: all bids are $\Vmax$, the CTRs are given by $\vec{\mu}_i$.

Let us focus on agents $1$ and $2$.
We claim that
    $\mathbb{E}_1[N_1]+\mathbb{E}_2[N_2] \geq \beta\,T^\lambda$,
where $\beta>0$ is a constant to be defined later. Suppose not. Fix all bids to be $\Vmax$. For each round $t$, consider event $S_t = \{ \A_t = 1\}$. Then by Claim~\ref{cl:LB-crux} we have
\begin{align*}
\bigl|\mathbb{P}_1[S_t] - \mathbb{P}_2[S_t]\bigr|
    \leq  \bigl|\mathbb{P}_0[S_t] - \mathbb{P}_1[S_t]\bigr|
        + \bigl|\mathbb{P}_0[S_t] - \mathbb{P}_2[S_t]\bigr|
    \leq O\left(\eps^2\right) \left(\mathbb{E}_1[N_1]+\mathbb{E}_2[N_2]\right)
    \leq \tfrac14
\end{align*}
for a sufficiently small $\beta$. Now, $\mathbb{P}_1[S_t]\geq  \tfrac12$ for at least $T/2$ rounds $t$. This is because otherwise on problem instance $\mathcal{I}_i$ regret would be
$R(T) \geq \Omega(\eps\, T \Vmax) = \Omega(\Vmax\, T^{1-\lambda/2})$, which contradicts the assumption $R(T) = O(\Vmax\,T^\gamma)$. Therefore
    $\mathbb{P}_2[S_t]\geq  \tfrac14$
for at least $T/2$ rounds $t$, hence on problem instance $\mathcal{I}_2$  regret is at least
    $\Omega(\eps\, T \Vmax)$,
contradiction. Claim proved.

Now without loss of generality let us assume that
    $\mathbb{E}_1[N_1] \geq \tfrac{\beta}{2}\,T^\lambda$.
Consider the problem instance in which CTRs given by $\vec{\mu}_1$, bid of agent $2$ is $\Vmax$, and all other bids are
    $\vmax({1-2\delta})/({1+2\epsilon})$.
It is easy to see that this problem instance has $\delta$-gap. Each time agent $1$ is selected, algorithm incurs regret $\Omega(\delta\Vmax)$. Thus the total regret is at least $\Omega(\delta N_1\,\Vmax) = \Omega(\delta\,\Vmax\,T^\lambda)$.
\end{proof}

%%%%%%%%%%%%%%%%%%%%%%%
%%%%%%%%%%%%%%%%%%%%%%

%%%%%%%%%%%%%%%%
\subsection{Relative entropy technique: proof of Claim~\ref{cl:LB-crux}}
\label{app:regret}

\newcommand{\WH}{\widehat{H}}
\newcommand{\Wh}{\widehat{h}}

We extend the relative entropy technique from~\cite{bandits-exp3}. All relevant facts about relative entropy are summarized in the theorem below. We will need the following definition: given a random variable $X$ on a probability space $(\Omega,\F,\mathbb{P})$, let $\mathbb{P}_X$ be the distribution of $X$, i.e. a measure on $\R$ defined by $\mathbb{P}_X(x) = \mathbb{P}[X=x] $.

\begin{theorem}[Some standard facts about relative entropy, e.g.~\cite{CoverThomas, Bobby-thesis,Bobby-LN}]~\\
\label{thm:KL-divergence}
Let $p$ and $q$ be two probability measures on a finite set $U$, and let $Y$ and $Z$ be functions on $U$. There exists a function $F(p;q|Y) : U \to \R$ with the following properties:
\begin{compactenum}[(i)]
\item $E_p\, F(p;q| Y) = E_p\, F(p;q| (Y,Z)) + E_p\, F(p_Z; q_Z| Y)$ \qquad({chain rule}),\label{item:chain-rule}
\item  $\bigl|p(U') - q(U')\bigr| \leq \sqrt{\tfrac{1}{2}\mathcal{D}(p \| q)}$ for any event $U'\subset U$, where $\mathcal{D}(p \| q) = E_p\,F(p;q|1)$ \label{item:bound-on-total-variation}
\item for each $x\in U$, if conditional on the event $\{Z=Z(x)\}$ $p$ coincides with $q$, then $F(p;q|Z)(x) = 0$. \label{item:zero-conditional-divergence}
\item for each $x\in U$, if conditional on the event $\{Z=Z(x)\}$ $p$ and $q$ are fair and $(\tfrac12+\eps)$-biased coins, respectively, then it is the case that $F(p;q|Z)(x) \leq 4 \eps^2$. \label{item:epsilon-squared-conditional-divergence}
\end{compactenum}
\begin{comment}
\begin{OneLiners}
\item[(i)] $E_p\, F(p;q| Y) = E_p\, F(p;q| (Y,Z)) + E_p\, F(p_Z; q_Z| Y)$
    ~~~(\emph{chain rule}),
\item[(ii)] $|p(U') - q(U')| \leq \sqrt{\tfrac12\, \mathcal{D}(p \| q)}$
    for any event $U'\subset U$, where
        $\mathcal{D}(p \| q) = E_p\,F(p;q|1)$
\item[(iii)] if conditional on the event $\{Z=z\}$, $p$ coincides with $q$, then $F(p;q|Z)(z) = 0$
\item[(iv)] if conditional on $\{Z=z\}$, $p$ and $q$ are fair and $(\tfrac12+\eps)$-biased coins, respectively, then it is the case that
     $F(p;q|Z)(z) = 8 \eps^2$.
\end{OneLiners}
\end{comment}
\end{theorem}

\begin{note}{Remark.}
This theorem summarizes several well-known facts about relative entropy, albeit in a somewhat non-standard notation. For the proofs, see~\cite{CoverThomas, Bobby-thesis,Bobby-LN}.
In the proofs, one defines $F = F(p;q|Y)$ as a function $F: U\to \R $ which is specified by
    $F(x)= \sum_{x'\in U} p(x'|U_x) \lg \tfrac{ p(x'|U_x)}{q(x'|U_x)} $,
where $U_x$ is the event $\{Y = Y(x)\}$.\footnote{We use the convention that
    $p(x) \log (p(x)/q(x))$
is 0 when $p(x)=0$, and $+\infty$ when $p(x)>0$ and $q(x)=0$.}
Note that the quantity
    $E_p\, F(p;q|1)$
is precisely the relative entropy (a.k.a. KL-divergence), commonly denoted $\mathcal{D}(p \| q)$, and
    $E_p\, F(p;q|Y)$
is the corresponding conditional relative entropy.
\end{note}

In what follows we use Theorem~\ref{thm:KL-divergence} to prove Claim~\ref{cl:LB-crux}. For simplicity we will prove~\refeq{eq:LB-crux} for $i=1$.

The \emph{history} up to round $t$ is
    $H_t = (h_1, h_2 \LDOTS h_t)$
where $h_s\in \{0,1\}$ is the click or no click event received by the algorithm at round $s$. Let $C_t$ be the indicator function of the event ``round $t$ is bid-independent". Define the \emph{bid-independent history} as
    $\WH_t = (\Wh_1, \Wh_2 \LDOTS \Wh_t)$,
where $\Wh_t = h_t C_t$.
For any exploration-separated deterministic allocation rule and each round $t$, the bid-independent history $\WH_{t-1}$ and the bids completely determine which arm is chosen in this round. Moreover, $\WH_{t-1}$ alone (without the bids) completely determines whether round $t$ is bid-independent, and if so, which arm is chosen in this round.

Recall the CTR vectors $\vec{\mu}_i$ as defined in Section~\ref{sec:regret}. Let $p$ and $q$ be the distributions induced on $\WH_T$ by $\vec{\mu}_0$ and $\vec{\mu}_1$, respectively. Let $p_t$ and $q_t$ be the distributions induced on $\Wh_t$ by $\vec{\mu}_0$ and $\vec{\mu}_1$, respectively. Let $\mathcal{H}_t$ the support of $\WH_t$, i.e. the set of all $t$-bit vectors.  In the forthcoming applications of Theorem~\ref{thm:KL-divergence}, the universe will be $U = \mathcal{H}_T$. By abuse of notation, we will treat $\WH_t$ as a projection $\mathcal{H}_T \to \mathcal{H}_t $, so that it can be considered a random variable under $p$ or $q$.

\begin{claim}\label{cl:chain-rule-application}
$\mathcal{D}(p \| q) = E_p\, F(p;q |\, \WH_t) \;+\;
    \sum_{s=1}^t E_p\, F(p_s;q_s|\, \WH_{s-1}) $
for any $t>1$.
\end{claim}

\begin{proof}
Use induction on $t\geq 0$ (set $\WH_0 = 1$). In order to obtain the claim for a given $t$ assuming that it holds for $t-1$, apply Theorem~\ref{thm:KL-divergence}(\ref{item:chain-rule}) with
    $Y = \WH_{t-1} $ and $Z = \Wh_t$.
\end{proof}

\begin{claim}
$F(p_t;q_t|\, \WH_{t-1}) \leq 4\eps^2\; C_t\; 1_{\{A_t = 1\}}$ for each round $t$.
\end{claim}
\begin{proof}
We are interested in the function
    $F = F(p_t;q_t|\, \WH_{t-1}):\, \mathcal{H}_T\to \R$.
Given $\WH_{t-1}$, one of the following three cases occurs:
\begin{compactitem}
%\begin{OneLiners}
\item round $t$ is not bid-independent. Then $\Wh_t=0$, hence $F(\cdot) = 0$ by Theorem~\ref{thm:KL-divergence}(\ref{item:zero-conditional-divergence}),
\item round $t$ is bid-independent and arm $1$ is not selected. Then $\Wh_t$ is distributed as a fair coin under both $p$ and $q$, so again $F(\cdot) = 0$.
\item round $t$ is bid-independent and arm $1$ is selected. Then  $F(\cdot) \leq 4\eps^2$ by Theorem~\ref{thm:KL-divergence}(\ref{item:epsilon-squared-conditional-divergence}). \qedhere
%\end{OneLiners}
\end{compactitem}
\end{proof}

Given the full bid-independent history $\WH_T$, $p$ and $q$ become (the same) point measure, so by Theorem~\ref{thm:KL-divergence}(\ref{item:zero-conditional-divergence})
    $E_p\, F(p;q |\, \WH_T) = 0$.
Therefore taking Claim~\ref{cl:chain-rule-application} with $t=T$ we obtain
\begin{align}\label{eq:LB2-divUB}
\mathcal{D}(p \| q)
    = \sum_{t=1}^T E_p\, F(p_t;q_t|\, \WH_{t-1})
    =  4\eps^2\; \sum_{t=1}^T E_p\, [C_t\; 1_{\{A_t = 1\}}]
    =  4\eps^2\; E_p[N_1].
\end{align}
For a given round $t$ and fixed bids, the allocation at round $t$ is completely determined by the bid-independent history $\WH_{t-1}$. Thus, we can treat
    $\{A_t \in S \}$
as an event in $\mathcal{H}_T$. Now~\refeq{eq:LB-crux} follows from ~\refeq{eq:LB2-divUB} via an application of Theorem~\ref{thm:KL-divergence}(\ref{item:bound-on-total-variation}) with $U' = \{A_t \in S \}$.

%%%%%%%%%%%%%%%%%%%%%%%%%%%%%%%%%%%
%%%%%%%%%%%%%%%%%%%%%%%%%%%%%%%%%%%

%%%%%%%%%%%%%%%%
\subsection{Lower bound for non-scalefree allocations}
\label{app:non-scalefree}

In this subsection we derive a regret lower bound for deterministic truthful mechanisms without assuming that the allocations are scale-free. In particular, for two agents there are no assumptions. This lower bound holds for any $k$ (the number of agents) assuming that the allocation satisfies IIA, but unlike the one in Theorem~\ref{thm:LB-main} it does not depend on $k$.

\begin{theorem}
\label{thm:LB-non-scalefree}
Consider the stochastic \problem\ with $k$ agents. Let $(\A,\mathcal{P})$ be a normalized truthful mechanism such that $\A$ is a non-degenerate deterministic allocation rule. Suppose $\A$ satisfies IIA. Then its regret is
   $R(T;\, \Vmax) = \Omega(\Vmax\, T^{2/3})$
for any sufficiently large $\Vmax$.
\end{theorem}

Let us sketch the proof. Fix an allocation \A. In Definition~\ref{def:structure}, if round $t$ is $(b,\rho)$ influential, for some \realization $\rho$ and bid vector $b$, an agent $i$ is called \emph{strongly influenced} by round $t$ if it is one of the two agents that are ``influenced" by round $t$ but is not the ``influencing agent" of round $t$. In particular, it holds that $\A(b,\rho,t)\neq i$. For each \realization $\rho$, round $t$ and agent $i$, if there exists a bid vector $b$ such that round $t$ is $(b,\rho)$-influential with strongly influenced agent $i$, then fix any one such $b$, and define
    $b^*_i = b^*_i(\rho,t) := \max_{j\neq i} b_j$.
Let us define
    $B^*_\A =\max_{\rho, t,i}\,b^*_i(\rho, t)$,
where the maximum is taken over all \realizations $\rho$, all rounds $t$, and all agents $i$. Let us say that round $t$ is \emph{$B^*$-free} from agent $i$ w.r.t \realization $\rho$, if for this \realization the following property holds: agent $i$ is not selected in round $t$ as long as each bid is at least $B^*$.

\begin{lemma}\label{lm:LB-nonSF}
In the setting of Theorem~\ref{thm:LB-non-scalefree}, for any \realization $\rho$, any influential round $t$ is $B^*_\A$-free from some agent w.r.t. $\rho$.
\end{lemma}

\begin{proof}
Fix \realization $\rho$. Since round $t$ is influential, for some bid profile $b$ and agent $i$ it is $(b,\rho)$-influential with a strongly influenced agent $i$. By definition of $b^*_i(\rho,t)$, without loss of generality each bid in $b$ (other than $i$'s bid) is at most
    $b^*_i(\rho,t) \leq B^*_\A$.
Then $\A(b,\rho,t)\neq i$, and round $t$ is $(b,\rho)$-secured from agent $i$.

Suppose round $t$ is not $B^*_\A$-free from agent $i$ w.r.t $\rho$. Then there exists a bid profile $b'$ in which each bid (other than $i$'s bid) is at least $B^*_\A$ such that
    $\A(b',\rho,t)= i$.
To derive a contradiction, let us transform $b$ to $b'$ by adjusting first the bid of agent $i$ and then bids of agents $j\neq i$ one agent at a time. Initially agent $i$ is not chosen in round $t$, and after the last step of this transformation agent $i$ is chosen. Thus it is chosen at some step, say when we adjust the bid of agent $i$ or some agent $j\neq i$. This {\em transfer of impression} to agent $i$ cannot happen when bid of agent $i$ is adjusted from $b_i$ to $b_i'$ (since round $t$ is $(b;\rho)$-secured from $i$), and it cannot happen when bid of agent $j\not=i$ is adjusted from $b_j$ to $b'_j\geq b_j$ (this is because, the transfer to $i$ cannot happen from $j$ because of pointwise-monotonicity and the transfer to $i$ cannot happen from $l\not=j$ because of IIA). This is a contradiction.
\end{proof}

\OMIT{ %%%%%%%
With Lemma~\ref{lm:LB-non-scalefree}, for $\Vmax> 2\,B^*_\A$ the proof of Theorem~\ref{thm:LB-main} carries over with minimal modifications. (All bids in that proof are either $\Vmax$ or $\Vmax/2$.)
} %%%

Let $T$ be the time horizon. Assume $\Vmax\geq 2 B^*_\A$. Let $N(\rho)$ be the number of influential rounds w.r.t \realization $\rho$. Let $N_i(\rho)$ be the number of influential rounds w.r.t. \realization $\rho$ that are $B^*_{\A}$-free from agent $i$ w.r.t.\ $\rho$. Then $N$ and the $N_i$'s are random variables in the probability space induced by the clicks. By Lemma~\ref{lm:LB-nonSF} we have that
    $\sum_i N_i(\rho)$ is at least the number of {\em influential rounds}.
As in Section~\ref{sec:regret}, let $\vec{\mu}_0$ be the vector of CTRs in which all CTRs are $\half$, and let
    $\E_0[\cdot]$
denote expectation w.r.t. $\vec{\mu}_0$.

Fix a constant $\beta>0$ to be specified later.  If
    $\E_0[N] \geq \beta k \,T^{2/3}$
then $\E_0[N_i] \geq \beta \,T^{2/3}$  for some agent $i$, so the allocation incurs expected regret
    $R(T; \Vmax) \geq \Omega(\Vmax\, T^{2/3})$
on any problem instance $\mathcal{J}_j$, $j\neq i$. (In this problem instance, CTRs given by $\vec{\mu}_0$, the bid of agent $j$ is $\Vmax$, and all other bids are $\Vmax/2$.) Now suppose
    $\E_0[N] \leq \beta k \,T^{2/3}$.
Then the desired regret bound follows by an argument very similar to the one in the last paragraph of the proof of Theorem~\ref{thm:LB-main}.

%%%%%%%%%%%%%%%%%%%%%%%%%%%%%%%%%%%
%%%%%%%%%%%%%%%%%%%%%%%%%%%%%%%%%%%
\subsection{Universally truthful randomized MAB mechanisms}
\label{app:universal}

Consider randomized mechanisms that are \emph{universally truthful}, i.e. truthful for each realization of the internal random seed. Our goal here is to extend the $\Omega(\Vmax\, T^{2/3})$ regret bounds for deterministic mechanisms to universally truthful randomized mechanisms, under relatively mild assumptions.

Note that lower bounds on regret for universally truthful MAB mechanisms do not immediately follow from those for deterministic truthful MAB mechanisms. To see this, consider a randomized MAB mechanism $\A$ that randomizes over some deterministic truthful mechanisms, each with regret at least $R$. Then for each deterministic mechanism $\A'$ in the support of $\A$ there is a problem instance on which $\A'$ has regret at least $R$; it could be a different problem instance for different $\A'$. Whereas to lower-bound the regret of $\A$ we need to provide one problem instance with high regret in expectation over \emph{all} $\A'$.

\OMIT{ %%%% A.S.: I don't think I will ever get myself to write out the proof ...
For the case of two agents, we obtain the desired result by extending the technique in Theorem~\ref{thm:LB-non-scalefree}:

\begin{theorem}\label{thm:univ-truthful-2}
Consider the \problem{} for two agents. Let $(\A,\mathcal{P})$ be a universally truthful mechanism such that $(\A,\mathcal{P})$ is universally normalized and $\A$ is universally non-degenerate. Then its expected regret is
   $\E \left[ R_{\A}(T;\Vmax) \right] = \Omega(\Vmax\, T^{2/3})$
for any sufficiently large $\Vmax$.
\end{theorem}

\begin{proof}
TO BE WRITTEN.
\end{proof}
We can extend this bound in Theorem~\ref{thm:univ-truthful-2} to $k>2$ agents (and, in fact, match the bounds in Theorems~\ref{thm:LB-main} and Theorem~\ref{thm:LB-delta}) if we assume that the mechanism is universally exploration-separated, i.e. that it randomizes over exploration-separated deterministic mechanisms.
}

We consider mechanisms that randomize over exploration-separated deterministic allocation rules. As per the discussion above, it does not suffice to quote Theorem~\ref{thm:LB-main}; instead, we need to extend its proof.

\newcommand{\RandAlloc}{\ensuremath{\mathcal{D}}}

\begin{lemma}
Consider the \problem. Let $\RandAlloc$ be a distribution over exploration-separated deterministic allocation rules. Then
    $$ \E_{\A\in \RandAlloc}\, \left[ R_{\A}(T;\Vmax) \right] =
    \Omega(\Vmax\, k^{1/3}\, T^{2/3}).
    $$
\end{lemma}

\begin{proof}
Recall that in the proof of Theorem~\ref{thm:LB-main} we define a family $\F$ of $2k$ problem instances, and show that if \A\ is an exploration-separated deterministic allocation rule, then on one of these instances its regret is ``high". In fact, we can extend this analysis to show that the regret is ``high", that is at least
    $R^* = \Omega(\Vmax\, k^{1/3}\, T^{2/3})$,
on an instance $\mathcal{I}\in\F$ chosen uniformly at random from \F; here regret is in expectation over the choice of $\mathcal{I}$.~\footnote{This extension requires but minor modifications to the proof of Theorem~\ref{thm:LB-main}. For instance, for the case $k\geq 3$ we argue that first, if $\E_0[N] > R$ then $\E_0[N_i] \leq \tfrac2k E_0[N]$ for at least $\tfrac{k}{2}$ agents $i$ (and so on), and if $\E_0[N] \leq R$ then (omitting some details) there are $\Omega(k)$ good agents $i$ such that $\E_0[N_i] \leq 2R/k$ (and so on).} Once this is proved, it follows that regret is $R^*/2$ for any \emph{distribution} over such \A, in expectation over both the choice of $\A$ and the choice of $\mathcal{I}$. Thus there exists a single (deterministic) instance $\mathcal{I}$ such that
$    \E_{\A\in \RandAlloc}\, \left[ R_{\A,\mathcal{I}}(T) \right] \geq R^*/2 $.
\end{proof}

Theorem~\ref{thm:LB-delta} can be extended similarly.

%\section{The naive MAB mechanism}
%\label{sec:naive}
%\begin{todo}
%[A \emph{short} proof of Observation~\ref{thm:naive-k}. This stuff is easy, let's not focus on it.]
%\end{todo}

\OMIT{%%%%%%%%%
To show that the worst-case lower bound on regret of Theorem~\ref{thm:LB-main} is tight
(up to a logarithmic factor) we next present the {\em naive MAB mechanism}, a simple truthful mechanism, with regret $R(T;\Vmax)= O(\Vmax\,  k^{1/3}\, T^{2/3}\, \log T)$.
} %%%%%%

\section{A matching upper bound}
\label{sec:naive}

Let us describe a very simple mechanism, called \emph{the naive MAB mechanism}, which matches the lower bound from Theorem~\ref{thm:LB-main} up to polylogarithmic factors (and also the lower bound from Theorem~\ref{thm:LB-delta}, for $\gamma=\lambda=\tfrac23$ and constant $\delta$).

Fix the number of agents  $k$, the time horizon $T$, and the bid vector $b$. The mechanism has two phases.  In the \emph{exploration phase}, each agent is selected for $T_0:= k^{-2/3}\,T^{2/3}(\log T)^{1/3}$ rounds, in a round robin fashion. Let $c_i$ be the number of clicks on agent $i$ in the exploration phase. In the \emph{exploitation phase}, an agent
    $i^* \in \argmax_i c_i b_i$
is chosen and selected in all remaining rounds. Payments are defined as follows: agent $i^*$ pays
    $\max_{i\in[k]\setminus \{i^*\}} c_i b_i/ c_{i^*}$ for every click she gets in exploitation phase,
and all others pay $0$. (Exploration rounds are free for every agent.) This completes the description of the mechanism.

\OMIT{ %%%%%%
We prove that the naive MAB mechanism is truthful, normalized, and has worst-case expected regret
    $R(T;\Vmax)= O(\Vmax\,  k^{1/3}\, T^{2/3}\, \log T)$,
see Appendix~\ref{app:naive} for details.
} %%%%%%%%
% This bound is tight for this mechanism since the mechanism can incur this much regret in exploration phase alone, e.g. if all CTRs are $\tfrac12$, the bid of agent $1$ is \Vmax, and all other bids are $\Vmax/2$.

\begin{lemma}\label{thm:naive-k}
Consider the stochastic \problem\ with $k$ agents. The naive mechanism is normalized,  truthful and has worst-case regret
    $R(T;\Vmax)= O(\Vmax\,  k^{1/3}\, T^{2/3}\, \log^{2/3} T)$.
\end{lemma}

\begin{proof}
The mechanism is truthful by a simple second-price argument.\footnote{Alternatively, one can use Theorem~\ref{thm:main-characterization} since all exploration rounds are bid-independent, and only exploration rounds are influential, and the payments are exactly as defined in Theorem~\ref{thm:Myerson-characterization}.}
Recall that $c_i$ is the number of clicks $i$ got in the exploration phase. Let $p_i=\max_{j\not=i}c_jb_j/c_i$ be the price paid (per click) by agent $i$ if she wins (all) rounds in exploitation phase. If $v_i\geq p_i$, then by bidding anything greater than $p_i$ agent $i$ gains $v_i-p_i$ utility each click irrespective of her bid, and bidding less than $v_i$, she gains $0$, so bidding $v_i$ is weakly dominant. Similarly, if $v_i<p_i$, then by bidding anything less than $p_i$ she gains $0$, while bidding $b_i>p_i$, she {\em loses} $b_i-p_i$ each click. So bidding $v_i$ is weakly dominant in this case too.

For the regret bound, let $(\ctr_1 \LDOTS \ctr_k )$ be the vector of CTRs, and let
    $\avgctr_i = c_i/T_0$
be the sample CTRs. By Chernoff bounds, for each agent $i$ we have
    $ \Pr \left[|\avgctr_i-\ctr_i| > r \right] \leq T^{-4},$
for $r=\sqrt{8\log (T)/T_0}$. If in a given run of the mechanism all estimates $\avgctr_i$ lie in the intervals specified above, call the run \emph{clean}. The expected regret from the runs that are not clean is at most $O(\Vmax)$, and can thus be ignored. From now on let us assume that the run is clean.

The regret in the exploration phase is at most
    $k\, T_0\, \vmax = O(\Vmax\,  k^{1/3}\, T^{2/3}\, \log^{1/3} T)$.
For the exploitation phase, let
    $ j = \argmax_i \ctr_i b_i$.
Then (since we assume that the run is clean) we have
\begin{align*}
(\ctr_{i^*}+r)\, b_{i^*}
    \geq \avgctr_{i^*}\, b_{i^*}
    \geq \avgctr_j\, b_j \geq (\ctr_j-r)\, b_j,
\end{align*}
which implies
    $ \ctr_j v_{j} - \ctr_{i^{*}} v_{i^*} \leq r(v_j + v_{i^*})\leq 2r\, \vmax$.
Therefore, the regret in exploitation phase is at most
    $2r\,\vmax\, T = O(\Vmax\,  k^{1/3}\, T^{2/3}\, \log^{2/3} T)$.
Therefore the total regret is as claimed.
\end{proof}

%%%%%%%%%%%%%%%%%%
\section{Randomized allocations and adversarially chosen clicks}
\label{sec:PSim}

In this section we discuss randomized allocations. We apply them to a version of the \problem\ in which clicks are generated adversarially.\footnote{We focus on the \emph{oblivious adversary} which (unlike the more difficult ``adaptive adversary") specifies all clicks in advance.}
The objective is to optimize the worst-case regret over all values
    $v = (v_1 \LDOTS v_k)$
such that $v_i \in [0,\Vmax] $ for each $i$, \emph{and all \realizations $\rho$}:
\begin{align}
R(T;v;\rho) &=
\left[ \textstyle{ \max_i  v_i \sum_{t=1}^T} \rho_i(t) \right] -
    \textstyle{\sum_{t=1}^T \sum_{i=1}^k}\, v_i\, \rho_i(t)\; \E\left[\A_i(v;\rho; t)\right]
        \label{eq:regret-def} \\
R(T;\Vmax) &= \max\{ R(T;v;\rho):\; \text{all \realizations $\rho$,
    all $v$ such that $v_i \in [0,\Vmax] $ for each $i$} \}. \nonumber
\end{align}
The first term in~\refeq{eq:regret-def} is the social welfare from the best time-invariant allocation, the second term is the social welfare generated by \A.

\OMIT{ %%% defns before Alex found a bug
Let us make a few definitions related to truthfulness. Recall that a mechanism is called \emph{weakly truthful} if for each \realization, it is truthful in expectation over its random seed. A randomized allocation is \emph{pointwise monotone} if for each \realization and each bid profile, increasing the bid of any one agent does not decrease the probability of this agent being allocated in any given round. For a set $S$ of rounds, an allocation is \emph{$S$-separated} if (i) for each \realization, the allocation in each round $t\in S$ (as a distribution over agents) does not depend on bids, (ii) the clicks from the rounds not in $S$ are discarded (not reported to the algorithm). An allocation is \emph{strongly separated} if before round $1$ it randomly chooses a set $S$ of rounds, without looking at the bids, and then runs a pointwise monotone $S$-separated allocation. Note that the choice of $S$ is independent of the clicks, by definition.
} %%%%%%

Let us make a few definitions related to truthfulness. Recall that a mechanism is called \emph{weakly truthful} if for each \realization, it is truthful in expectation over its random seed. A randomized allocation is \emph{pointwise monotone} if for each \realization and each bid profile, increasing the bid of any one agent does not decrease the probability of this agent being allocated in any given round. For a set $S$ of rounds and a function $\sigma:S\to \{\text{agents}\}$, an allocation is \emph{$(S,\sigma)$-separated} if (i) it coincides with $\sigma$ on $S$, (ii) the clicks from the rounds not in $S$ are discarded (not reported to the algorithm). An allocation is \emph{strongly separated} if before round $1$, without looking at the bids, it randomly chooses a set $S$ of rounds and a function $\sigma:S\to \{\text{agents}\}$, and then runs a pointwise monotone $(S,\sigma)$-separated allocation. Note that the choice of $S$ and $\sigma$ is independent of the clicks, by definition.

\newcommand{\PSim}{{\sc PSim}}
\newcommand{\DaniHayes}{{\sc DaniHayes}}

We obtain a structural result: for any (randomized) strongly separated allocation rule $\A$ there exists
% a payment rule which results in
a mechanism that is normalized and weakly truthful.

\begin{lemma}
\label{lm:weakly-truthful}
Consider the \problem. Let $\A$ be a (randomized) strongly separated allocation rule. Then there exists a payment rule $\price$ such that the resulting mechanism $(\A,\price)$ is normalized and weakly truthful.
\end{lemma}

We consider \PSim~\cite{Bobby-stoc04-journal-version,Robert-Kleinberg-Lecture-8}, a randomized MAB algorithm from the literature which we here interpret as an MAB allocation rule. It follows from~\cite{Bobby-stoc04-journal-version,Robert-Kleinberg-Lecture-8}, that \PSim{} has strong regret guarantees for the adversarial \problem: it obtains regret
    $R(T,\Vmax) = O(\Vmax\, k^{1/3}\, (\log k)^{1/3}\, T^{2/3} )$.
In Section~\ref{app:PSim} we state \PSim{} and show that it is strongly separated. Thus, we obtain the following result.

% We state the above results in the form of the following corollary.
\begin{theorem}\label{cor:weakly-truthful}
There exists a weakly truthful normalized mechanism for the adversarial MAB problem (against oblivious adversary) whose regret grows as
    $\bigo( (k\log k)^{1/3}\cdot T^{2/3}\cdot \vmax)$.
\end{theorem}

\begin{note}{Remark.}
For the adversarial MAB problem (i.e., without the restriction of truthfulness), the regret bound can be improved to $\tilde{O}(\sqrt{kT}\cdot\Vmax)$~\cite{bandits-exp3, Bubeck-colt09}. However, the algorithms that achieve this bound do not immediately yield MAB allocation rules that are strongly separated. It is an open question whether the regret bound in Corollary~\ref{cor:weakly-truthful} can be improved.
\end{note}

\begin{proofof}{Lemma~\ref{lm:weakly-truthful}}
Throughout the proof, let us fix a \realization $\rho$, time horizon $T$, bid vector $b$, and agent $i$. We will consider the payment of agent $i$. We will vary the bid of agent $i$ on the interval $[0, b_i]$; the bids $b_{-i}$ of all other agents always stay the same.

Let $c_i(x)$ be the number of clicks received by agent $i$ given that her bid is $x$. Then by (the appropriate version of) Theorem~\ref{thm:Myerson-characterization} the payment of agent $i$ must be $\price_i(b)$ such that
\begin{align}\label{eq:app-PSim-Myerson}
\E_{\A}[\price_i(b)] = \E_\A \left[
    b_i\, c_i(b_i) - \textstyle{\int_{x=0}^{b_i}}\, c_i(x) \d x
    \right],
\end{align}
where the expectation is taken over the internal randomness in the algorithm.

Recall that initially $\A$ randomly selects, without looking at the bids, a set $S$ of rounds and a function $\sigma:S\to\{\text{agents}\}$, and then runs some pointwise monotone $(S,\sigma)$-separated allocation $\A^{(S,\sigma)}$.  In what follows, let us fix $S$ and $\sigma$, and denote $\A^* = \A^{(S,\sigma)}$. We will refer to the rounds in $S$ as \emph{exploration rounds}, and to the rounds not in $S$ as \emph{exploitation rounds}. Let
    $\gamma^*_i(x,t)$
be the probability that algorithm $\A^*$ allocates agent $i$ in round $t$ given that agent $i$ bids $x$. Note that for fixed value of internal random seed of $\A^*$ this probability can only depend on the clicks observed in exploration rounds, which are known to the mechanism. Therefore, abstracting away the computational issues, we can assume that it is known to the mechanism. Define the payment rule as follows: in each exploitation round $t$ in which agent $i$ is chosen and clicked, charge
\begin{align}\label{eq:app-PSim-payment}
\price^*_i(b,t) = b_i- \frac{1}{\gamma^*_i(b_i,t)}
        \int_{0}^{b_i}\gamma^*_i(x,t) \,dx.
\end{align}
Then the total payment assigned to agent $i$ is
\begin{align}\label{eq:app-PSim-payment-total}
\price^*_i(b)
    &= \textstyle{ \sum_{t \not\in S} }\;
        \rho_i(t)\; \A^*_i(b;\rho; t)\; \price^*_i(b,t).
\end{align}

Since allocation $\A^*$ is pointwise monotone, the probability $\gamma^*_i(x,t)$ is non-decreasing in $x$. Therefore
    $\price^*_i(b,t)\in [0, b_i]$
for each round $t$. It follows that the mechanism is normalized (for any realization of the random seed of allocation \A).

It remains to check that the payment rule~\refeq{eq:app-PSim-payment} results in~\refeq{eq:app-PSim-Myerson}. Let
    $c^*_i(x)$
be the number of clicks allocated to agent $i$ by allocation $\A^*$ given that her bid is $x$. Let
    $c^{\text{expl}}_i(x)$
be the corresponding number of clicks in exploitation rounds only. Since $\A^*$ is $(S,\sigma)$-separated, we have
\begin{align}\label{eq:app-PSim-separated}
\E[c^*_i(x) - c^{\text{expl}}_i(x) ]
    = \textstyle{\sum_{t\in S}}\; \rho_{\sigma(t)}(t)\;
    = \text{const}(x).
\end{align}
Taking expectations in~\refeq{eq:app-PSim-payment-total} over the random seed of $\A_S$ and using~\refeq{eq:app-PSim-separated}, we obtain
\begin{align*}
\E [\price^*_i(b)]
    &= \textstyle{ \sum_{t \not\in S} }\;
        \rho_i(t)\; \gamma^*_i(b_i,t)\; \price^*_i(b,t) \\
    &= \textstyle{ \sum_{t \not\in S} }\;
       \rho_i(t)\, \left[ b_i\, \gamma^*_i(b_i,t) -
        \textstyle{\int_0^{b_i}}\, \gamma^*_i(x,t)\, dx \right] \\
    &= b_i \left[ \textstyle{ \sum_{t \not\in S} }\;
        \rho_i(t)\; \gamma^*_i(b_i,t) \right]
        - \textstyle{\int_0^{b_i}}\,
        \left[
            \textstyle{ \sum_{t \not\in S} }\; \rho_i(t)\,\gamma^*_i(x,t)
        \right] \, dx \\
    &= b_i\, \E\,[c^{\text{expl}}_i(b_i)] - \textstyle{\int_0^{b_i}}\, \E[ c^{\text{expl}}_i(x)]\, dx \\
    &= \E \left[
        b_i\, c^*_i(b_i) - \textstyle{\int_0^{b_i}}\, c^*_i(x)\, dx
        \right].
\end{align*}
Finally, taking expectations over the choice of $S$ and $\sigma$, we obtain~\refeq{eq:app-PSim-Myerson}.
\end{proofof}

\subsection{Algorithm \PSim{} is strongly separated}
\label{app:PSim}

In this subsection we interpret \PSim~\cite{Bobby-stoc04-journal-version,Robert-Kleinberg-Lecture-8} as an MAB allocation rule and show that it is strongly separated (which implies Theorem~\ref{cor:weakly-truthful}). For the sake of completeness, we present \PSim{} below. As usual, $k$ denotes the number of agents; let $[k]$ denote the set of agents.

\smallskip
\begin{compactenum}
\item[\bf Input:] Time horizon $T$, bid vector $b$. Let $\vmax=\max_{i}b_{i}$.
\item[\bf Output:] For each round $t\leq T$, a distribution on $[k]$.
\item Divide the time horizon into $P$ phases of $T/P$ consecutive rounds each.
\begin{comment}
%\item Pick number of phases $P$ and $Q=T/P$. Divide the time horizon into $P$ phase of length $Q$ each. For phase $p=1,2,\dots,P$, let $\phi_p$ denote the sequence of rounds in phase $p$. That is
	$$\phi_p=\{(p-1)Q+1,\; (p-1)Q+2  \LDOTS\ pQ\}.$$
\end{comment}
%\item For each phase $\phi_p$, pick a random one-to-one function $f_p:K\to \phi_p$, and call $f_p(K)$ as exploration rounds in phase $p$.
\item From rounds of each phase $p$, pick without replacement $k$ rounds at random (called the {\em exploration rounds}) and assign them randomly to $k$ arms. Let $S$ denote the set of all exploration rounds (of all phases). Let $f:S\to [k]$ be the function which tells which arm is assigned to an exploration round in $S$. The rounds in $[T]\setminus S$ are called the exploitation rounds.
%and rounds $S_i:=\cup_{p}f_p(i)$ as exploration rounds for agent $i$.
\item Let $w_i(0)=1$ for all $i\in [k]$.
\item For each phase $p=1,2,\dots,P$
	\begin{compactenum}
	\item For each round $t$ in phase $p$	
		\begin{compactenum}
		\item If $t\in S$ and $f(t)=i$, then define the distribution $\gamma(b;t;S,f)$ such that $\gamma_i(b;t;S,f)=1$. Pick an agent according to this distribution (equivalently, pick agent $i$), observe the click $\rho_i(t)$, and update $w_i(p)$ multiplicatively,
    $$w_i(p)=w_i(p-1)\cdot (1+\epsilon)^{\rho_i(t) b_i/\vmax}.$$
		\item If $t\not\in S$, then define the distribution $\gamma(b;t;S,f)$ such that $\gamma_i(b;t;S,f)=\frac{w_i(p-1)}{\sum_{j}w_j(p-1)}$. Pick an agent according to $\gamma(b;t;S,f)$, observe the feedback, and discard the feedback.
		\end{compactenum}
	\end{compactenum}
\end{compactenum}

\xhdr{Regret.} If we pick the values $\epsilon=(k\log k/T)^{1/3}$ and $P=(\log k)^{1/3}(T/k)^{2/3}$, then the regret of $\psim$ is bounded by $\bigo((k\log k)^{1/3}T^{2/3}\vmax)$ against any oblivious adversary (see \cite{Bobby-stoc04-journal-version,Robert-Kleinberg-Lecture-8}).

\begin{claim}
$\psim$ is strongly-separated.
\end{claim}

\begin{proof}
It is clear from the structure of $\psim$ above that it chooses a set $S$ of exploration rounds and a function $f:S\to [k]$ in the beginning without looking at the bids and then runs an $(S,f)$-separated allocation. We need to prove that the $(S,f)$-separated allocation is pointwise monotone. For this we need prove that the probability $\gamma_i(b;t; S,f)$ is monotone in the bid of agent $i$, where $\gamma_i(b;t;S,f)$ denotes the probability of picking agent $i$ in round $t$ when bids are $b$ given the choice of $S$ and $f$. If $t\in S$, the $\gamma_i(b;t;S,f)$ is independent of bids, and hence is monotone in $b_i$. Let $t\not\in S$ and $t$ is a round in phase $p$. Let us denote by $f^{-1}(i,p)$ the (unique) exploration round in phase $p$ assigned to agent $i$. We then have
\begin{align*}
\gamma_i(b;t; S,f) = (1+\epsilon)^{\frac{b_i}{\vmax}\sum_{q=1}^{p-1}\rho_i(f^{-1}(i,q))} \biggm/ \sum_{j}(1+\epsilon)^{\frac{b_j}{\vmax}\sum_{q=1}^{p-1}\rho_j(f^{-1}(j,q))}.
\end{align*}
We split the denominator into the term for agent $i$ and all other terms. It is then not hard to see that this is a non-decreasing function of $b_i$.
\end{proof}

\section{Truthfulness in expectation over CTRs}
\label{sec:expectation}

% notation for "Myerson polynomial"
\newcommand{\PMy}{\ensuremath{\mathcal{P}^{\text{M}}}}

% notation for "exploration allocation rule"
\newcommand{\Aexpl}{\ensuremath{\A_{\text{expl}}}}

We consider the stochastic \problem\ under a more relaxed notion of truthfulness: truthfulness \emph{in expectation}, where for each vector of CTRs the expectation is taken over clicks (and the internal randomness in the mechanism, if the latter is not deterministic).\footnote{
\emph{Normalized-in-expectation} and \emph{monotone-in-expectation} properties are defined similarly. An allocation rule is \emph{monotone in expectation} if for each agent $i$ and fixed bid profile $b_{-i}$, the corresponding expected click-allocation is a non-decreasing function of $b_i$. A mechanism is \emph{normalized in expectation} if in expectation each agent is charged an amount between $0$ and her bid for each click she receives. In both cases, the expectation is taken over the clicks and possibly the allocation's random seed.} We show that any MAB allocation $\A^*$ that is monotone in expectation, can be converted to an MAB mechanism that is truthful in expectation and normalized in expectation, with minor changes and a very minor increase in regret.  As discussed in the Introduction, this result rules out a natural lower-bounding approach.

\begin{note}{Remark.}
The follow-up work~\cite{Transform-ec10} has established that there exist MAB allocations that are monotone in expectation whose regret matches the optimal upper bounds for MAB \emph{algorithms}. In fact,~\cite{Transform-ec10} defined a rather natural class of ``well-formed MAB algorithms" that, e.g., includes (a version of) algorithm \UCB~\cite{bandits-ucb1}, and proved that any algorithm in this class gives rise to a monotone-in-expectation MAB allocation.
\end{note}

\OMIT{Furthermore, we show that there exist MAB allocations that are monotone in expectation whose regret matches the optimal upper bounds for MAB \emph{algorithms}. The conclusion is that in order to obtain any non-trivial lower bounds on regret and (essentially) any non-trivial structural results, one needs to assume that a mechanism is ex-post normalized, at least in some approximate sense.}

We will show that for any allocation $\A^*$ that is monotone in expectation, any time horizon $T$, and any parameter $\gamma\in(0,1)$ there exists a mechanism $(\A,\mathcal{P})$ such that the mechanism is truthful in expectation and normalized in expectation, and allocation $\A$ initially makes a random choice between $\A^*$ and some other allocation, choosing $\A^*$ with probability at least $\gamma$. We call such allocation $\A$ a \emph{$\gamma$-approximation} of $\A^*$. Clearly, on any problem instance we have
    $R_\A(T) \leq \gamma\,R_{\A^*}(T) + (1-\gamma) T $.
The extra additive factor of $(1-\gamma) T$ is not significant if e.g.
    $\gamma = 1-\tfrac{1}{T}$.
The problem with this mechanism is that it is not ex-post normalized; moreover, in some \realizations payments may be very large in absolute value.

\OMIT{We provide a (rather weak) upper bound on the \emph{PPC-to-bid ratio}, the ratio between the absolute value of the payment-per-click and the bid.}

\begin{theorem}\label{thm:in-expectation}
Consider the stochastic \problem\ with $k$ agents and a fixed time horizon $T$. For each $\gamma\in(0,1)$ and each allocation rule $\A^*$ that is monotone in expectation, there exists a mechanism $(\A, \mathcal{P})$ such that \A\ is a $\gamma$-approximation of $\A^*$, and the mechanism is truthful in expectation and normalized in expectation.
\end{theorem}

\OMIT{For each \realization, the PPC-to-bid ratio of the mechanism is upper-bounded by a function of $k$, $T$ and $\gamma$.}

\begin{note}{Remark.}
The key idea is to view the Myerson payments (see Theorem~\ref{thm:Myerson-characterization}) as multivariate polynomials over the CTRs, and argue that any such polynomial can be ``implemented" by a suitable payment rule. The payment rule $\mathcal{P}$ will be well-defined as a mapping from histories to numbers; we do not make any claims on the efficient computability thereof.
\end{note}

\OMIT{ %%%%%%%%% false claim!!
For the sake of completeness, we provide a concrete algorithm which one could plug into Theorem~\ref{thm:in-expectation} and obtain improved (and in fact, best possible) regret guarantees.

\begin{proposition}\label{thm:monotone-in-expectation}
Consider the stochastic \problem\ with $k$ agents and a fixed time horizon $T$. There exists an allocation rule \A\ that is monotone in expectation, whose regret is
   $R(T;\, \Vmax) = O(\Vmax\, \sqrt{kT \log T})$
in the worst case, and
    $R_\delta(T;\, \Vmax) = O(\Vmax\, \tfrac{k}{\delta} \log T)$
on the $\delta$-gap instances.
\end{proposition}

\begin{proof}[Proof Sketch]
For simplicity, assume $\Vmax=1$. Let $r_0 = \sqrt{8\log(T)/T}$. Consider the following simple allocation.  Initially, each agent is \emph{active}. In each phase, select each active agent once, in a round-robin fashion. After the phase, (permanently) de-activate each agent whose \emph{sample product} (sample average times the bid) is more than $r_0$ below that of some other active agent. This completes the description of the allocation.

This allocation is based on a well-known (perhaps folklore) MAB algorithm. The regret bounds are proved along the lines of those in~\cite{bandits-ucb1}. The crucial observations are that with a very high probability the optimal agent is never de-activated, and that that each sub-optimal agent $i$ is selected at most $O(\Delta_i^{-2}\, \log T)$ times, where $\Delta_i$ is the difference between her product (CTR times the bid) and the maximal one.

The allocation is monotone in expectation because increasing the bid of a given agent cannot cause this agent to be de-activated later.
\end{proof}
} %%%%%%%%%%%

%%%%%%%%%%%%%%%
%\subsection{Proof of Theorem~\ref{thm:in-expectation}}

\begin{proof}
Let \Aexpl\  be the allocation rule where in each round an agent is chosen independently and uniformly at random. Allocation \A\ is defined as follows:  use $\A^*$ with probability $\gamma$; otherwise use \Aexpl. Fix an instance $(b,\mu)$ of the stochastic \problem, where
    $b = (b_1 \LDOTS b_k)$ and $\mu = (\mu_1\LDOTS \mu_k)$
are vectors of bids and CTRs, respectively. Let $C_i = C_i(b_i; b_{-i})$ be the expected number of clicks for agent $i$ under the original allocation $\A^*$. Then by Myerson~\cite{Myerson} the expected payment of agent $i$ must be
\begin{align}\label{eq:in-expectation-PMy}
 \PMy_i =
   \gamma\left[
        b_i\, C_i(b_i; b_{-i}) - \textstyle{\int_0^{b_i}} C_i(x; b_{-i})\, dx
        \right].
\end{align}

\noindent We treat the expected payment as a multivariate polynomial over
    $\mu_1 \LDOTS \mu_k$.

\OMIT{It is essential (given the way we define $\mathcal{P}$) to show that this polynomial has degree $\leq T$.}

\OMIT{Also, for upper-bounding the PPC-to-bid ratio it is useful to upper-bound the absolute value of the coefficients.}

\begin{claim}\label{cl:in-expectation-PMy}
$\PMy_i$ is a polynomial of degree $\leq T$ in variables $\mu_1 \LDOTS \mu_k$.
\OMIT{The absolute value of the coefficients of this polynomial (after the similar terms are added) is at most
    $b_i\,T\,(4k)^T$.}
\end{claim}

% notation for the set of all polynomials and monomials resp.
\newcommand{\POLY}{\ensuremath{\mathtt{poly}}}
\newcommand{\MONO}{\ensuremath{\mathtt{mono}}}

\begin{proof}
Fix the bid profile. Let $X_t$ be allocation of algorithm $\A^*$. Let $\POLY(T)$ be the set of all polynomials over $\mu_1 \LDOTS \mu_k$ of degree at most $T$. Consider a fixed history
    $h = (x_1, y_1; \;\ldots\; ;x_T,y_T)$,
and let $h^t$ be the corresponding history up to (and including) round $t$. Then
\begin{align}
\mathbb{P}[h] &= \textstyle{\prod_{t=1}^T}
    \Pr[X_t = x_t \given h^{t-1}]\;\; \mu_{x_t}^{y_t}\, (1-\mu_{x_t})^{1-y_t}
        \in \POLY(T)
        \label{eq:in-expectation-prob-h} \\
C_i(b_i; b_{-i}) &=
    \textstyle{\sum_{h\in \mathcal{H}}}\,
        \mathbb{P}[h]\; \#\text{clicks}_i(h)\,
        \in \POLY(T). \label{eq:in-expectation-Ci}
\end{align}
Therefore $\PMy_i\in \POLY(T)$, since one can take an integral in~\refeq{eq:in-expectation-PMy} separately over the coefficient of each monomial of $C_i(x; b_{-i})$.
\OMIT{%%%
For a polynomial $Q$, let $\| Q\|_\infty$ be the maximal absolute value of its coefficients. To upper-bound $\| \PMy_i \|_\infty$, note that by~\refeq{eq:in-expectation-prob-h}, a (crude) upper bound on
$\| \mathbb{P}[h] \|_\infty$ is $2^T$, so
\begin{align*}
\| \PMy_i \|_\infty
    \leq b_i\; \| C_i(b_i; b_{-i}) \|_\infty
    \leq b_i\cdot T \cdot |\mathcal{H}|\cdot \max_h \|\mathbb{P}[h] \|_\infty
    \leq b_i\, T\,(4k)^T. \qquad \qedhere
\end{align*}
}%%% \OMIT
\end{proof}

Fix time horizon $T$. For a given run of an allocation rule, the \emph{history} is defined as
    $h = (x_1, y_1; \;\ldots\; ;x_T,y_T)$,
where $x_t$ is the allocation in round $t$, and $y_t\in\{0,1\}$ is the corresponding click. Let $\mathcal{H}$ be the set of all possible histories.

Our payment rule $\mathcal{P}$ is a deterministic function of history. For each agent $i$, we define the payment
    $\mathcal{P}_i  = \mathcal{P}_i(h)$
for each history $h$ such that
    $E_h[ \mathcal{P}_i(h)] = \PMy_i$
for any choice of CTRs, and hence
    $E_h[\mathcal{P}_i(h)] \equiv \PMy_i$,
where $\equiv$ denotes an equality between polynomials over
    $\mu_1 \LDOTS \mu_k$.

Fix the bid vector and fix agent $i$. We define the payment $\mathcal{P}_i$ as follows.  Charge nothing if allocation $\A^*$ is used. If allocation \Aexpl\ is used, charge \emph{per monomial}. Specifically, let $\MONO(T)$ be the set of all monomials  over $\mu_1 \LDOTS \mu_k$ of degree at most $T$. For each monomial $Q\in\MONO(T)$  we define a subset of \emph{relevant histories}
    $\mathcal{H}_i(Q) \subset \mathcal{H}$.
(We defer the definition till later in the proof.)
For a given history $h\in \mathcal{H}$ we charge a (possibly negative) amount
\begin{align}\label{eq:in-expectation-payment}
\mathcal{P}_i(h) = \tfrac{1}{1-\gamma}\;
        \textstyle{\sum_{Q\in \MONO(T):\, h\in \mathcal{H}_i(Q)}}\;
             k^{\deg(Q)}\; \PMy_i(Q),
\end{align}
where $\deg(Q)$ is the degree of $Q$, and $\PMy_i(Q)$ is the coefficient of $Q$ in $\PMy_i$. Let $\mathbb{P}_\text{expl}$ be the distribution on histories induced by  \Aexpl. Then the expected payment is
$$E_h [\mathcal{P}_i(h)] =
    \textstyle{\sum_{Q\in\MONO(T)}}\; k^{\deg(Q)}\;
        \mathbb{P}_\text{expl}[\mathcal{H}_i(Q)]\; \PMy_i(Q).
$$
Therefore in order to guarantee that
    $E_h[ \mathcal{P}_i(h)] \equiv  \PMy_i$
it suffices to choose $\mathcal{H}_i(Q)$ for each $Q$ so that
\begin{align}\label{eq:in-expectation-HQ}
k^{\deg(Q)}\;\mathbb{P}_\text{expl}[\mathcal{H}_i(Q)] \equiv Q.
\end{align}
Consider a monomial
    $Q = \mu_1^{\alpha_1}\,\ldots \, \mu_k^{\alpha_k}$.
Let $\mathcal{H}_i(Q)$ consist of all histories such that first agent $1$ is selected $\alpha_1$ times in a row, and clicked every time, then agent $2$ is selected $\alpha_2$ times in a row, and clicked every time, and so on till agent $k$. In the remaining
    $T-\deg(Q)$
rounds, any agent can be chosen, and any outcome (click or no click) can be received. It is clear that~\refeq{eq:in-expectation-HQ} holds.
\end{proof}

\OMIT{ %%%%%%%%
Note that a given history can be relevant to at most one monomial of a given degree, so it can be relevant to at most $T$ monomials. Since we know from Claim~\ref{cl:in-expectation-PMy} that
    $\|\PMy_i(Q)\|_\infty \leq b_i\,T\,(4k)^T$,
it follows from by~\refeq{eq:in-expectation-payment} that
    $|\mathcal{P}_i(h)| \leq \tfrac{1}{1-\gamma}\; b_i\,T^2\,(2k)^{2T}$.
} %%%%%% \OMIT

\section{Open questions}
\label{sec:questions}

Despite the exciting developments in the follow-up work~\cite{Transform-ec10,SingleCall-ec12,Gatti-ec12,Parkes-netecon12} (discussed in Section~\ref{sec:followup}), MAB mechanisms are not well-understood. Below is a snapshot of the open questions, current as of this writing.

\xhdr{Impossibility results for deterministic MAB mechanisms.}

\begin{enumerate}

\item For deterministic MAB mechanisms with $k>2$ agents, is it possible to
obtain lower bounds on regret for weakly separated MAB allocation rules, without assuming IIA?

\item  We conjecture that the ``informational obstacle'' -- insufficient observable information to compute payments -- can be meaningfully extended to a very general class of mechanisms in which an allocation rule interacts with the environment. As mentioned in Section~\ref{sec:followup}, the follow-up work~\cite{SingleCall-ec12,Parkes-netecon12} suggested settings other than MAB mechanisms in which this obstacle arises. To conclude that the ``informational obstacle'' is prominent in a given setting, one needs to prove that unrestricted payment computation makes truthful mechanisms strictly more powerful.

\item Surprisingly, we still do not understand the limitations of deterministic truthful-in-expectation mechanisms. While, according to~\cite{Transform-ec10}, there exist regret-optimal MAB allocation rules that are deterministic and monotone-in-expectation (e.g., the allocation rule based on \UCB), it is not clear whether any such allocation rule can be extended to a \emph{deterministic} truthful-in-expectation MAB mechanism.

\item It would be interesting to analyze a slightly more permissive model in which an MAB mechanism can decide to ``skip'' a round without displaying an ad. In particular, in such model we could trivially extend the lower bounds on regret from the special case of $k=2$ agents to $k>2$ agents. However, our negative results for two agents do not immediately extend to this new model, and moreover the structural results for $k>2$ agents do not immediately follow either.

\end{enumerate}

\xhdr{Randomized MAB mechanisms.}

\begin{enumerate}
\item Recall that the ``BKS reduction'' from Babaioff, Kleinberg and Slivkins~\cite{Transform-ec10} exhibits a tradeoff between variance in payments and loss in performance. Since the variance in payments can be very high, optimizing this tradeoff is crucial.

    This question is \emph{not} resolved by the worst-case optimality result in Wilkens and Sivan~\cite{SingleCall-ec12}. While no other reduction can achieve a better tradeoff for all monotone MAB allocation rules simultaneously, the result in~\cite{SingleCall-ec12} does not rule out a reduction with better tradeoff for \emph{some} monotone MAB allocation rules, and therefore it does not rule out an MAB mechanism with better tradeoff. Furthermore, it is possible that an MAB mechanism with optimal tradeoff cannot be represented as a reduction from a regret-optimal allocation rule, in which case results about reductions simply do not apply.

\item  Consider weakly truthful MAB mechanisms in the setting with adversarially chosen clicks.%
\footnote{Recall that an MAB mechanism is weakly truthful if for each click realization, it is truthful in expectation over its random seed. Weakly monotone MAB allocation rules are defined similarly.} The weakly truthful MAB mechanism in the present paper achieves regret $\tilde{O}(k^{1/3}\,T^{2/3})$, whereas the best known MAB algorithms achieve regret $O(\sqrt{kT})$~\cite{bandits-exp3,Bubeck-colt09}. It is not clear what should be the tight regret bound. In particular,
neither our reduction in Section~\ref{sec:PSim} nor the BKS reduction from~\cite{Transform-ec10} immediately apply to the algorithms in~\cite{bandits-exp3,Bubeck-colt09}.

\item More generally, as discussed in Section~\ref{sec:related-work}, pay-per-click ad auctions motivate many other versions of the \problem, corresponding to the various MAB settings studied in the literature. For every such version one could compare the performance of weakly truthful MAB mechanisms with that of the best MAB algorithms. The positive direction here reduces (using the BKS reduction) to designing weakly monotone MAB allocations. This type of question is a new angle in the MAB literature, see~\cite{MonotoneMAB-colt11} for a self-contained account.

\end{enumerate}

\xhdr{Multi-slot MAB mechanisms:} pay-per-click auctions with multiple ad slots and unknown CTRs.

\begin{enumerate}

\item Intuitively it seems that the negative results from this paper should extend to the setting with two or more ad slots. However, the precise characterization results and regret bounds remain elusive. Also, such results would probably depend on the specific multi-slot model, i.e. on on how clicks in different slots are correlated, and how CTRs of the same ad in different slots are related to one another.

\item Recall that Gatti, Lazaric and Trovo~\cite{Gatti-ec12} provide truthful multi-slot MAB mechanisms based on the simple MAB mechanism presented in this paper and (independently) in Devanur and Kakade~\cite{DevanurK09}. It remains to be seen if one can obtain weakly truthful mechanisms with better regret, e.g. using a more efficient multi-slot MAB algorithm with an extension of the BKS reduction. Note that even the algorithmic (i.e., non-strategic) version of multi-slot MAB is not fully understood.

\end{enumerate}

\section*{Acknowledgements}

We thank Jason Hartline, Robert Kleinberg and Ilya Segal for helpful discussions.

% ############ BIBLIOGRAPHY ##############

\begin{small}
\bibliographystyle{plain}
\bibliography{bib-abbrv,bib-bandits,bib-AGT,bib-slivkins}
\end{small}

\appendix
%%%%%%%%%%%%%%
\section{Proof of Lemma~\ref{lem:scalefree-and-iia-implies-equivalent-conditions}}
\label{sec:IIA}

In this section we present the full proof of Lemma~\ref{lem:scalefree-and-iia-implies-equivalent-conditions}.
Recall that
the ``only if'' direction is a consequence of Observation~\ref{obs:weakly-separated}. We focus on the ``if'' direction.

For bid profile $b$, \realization $\rho$, agent $l$ and round $t$, the tuple $(b; \rho; l; t)$ is called an \emph{influence-tuple} if round $t$ is $(b,\rho)$-influential with influencing agent $l$. Suppose allocation $\A$ is weakly separated but not exploration-separated. Then there is a \emph{counterexample}: an influence-tuple $(b; \rho; l; t)$ such that round $t$ is not bid-independent w.r.t. \realization $\rho$.  We prove that such counterexample can occur only if $b_l\in S_l(b_{-l})$, for some finite set $S_l(b_{-l})\subset \R$ that depends only on $b_{-l}$.

\begin{proposition}\label{prop:crux}
Let \A\ be as in Lemma~\ref{lem:scalefree-and-iia-implies-equivalent-conditions}. Assume \A\ is weakly separated. Then for each agent $l$ and each bid profile $b_{-l}$ there exists a finite set $S_l(b_{-l})\subset \R$ with the following property: for each counterexample
    $(b_l, b_{-l}; \rho; l; t)$ it is the case that $b_l\in S_l(b_{-l})$.
\end{proposition}

\OMIT{ %%%%%%%
for each counterexample $(b; \rho; l; t)$ it is the case that $b_l$ belongs to a finite set $S_l(b_{-l})\subset \R$ that depends only on $b_{-l}$.
}

Once this proposition is proved, we obtain a contradiction with the non-degeneracy of \A. Indeed, suppose $(b; \rho; l; t)$ is a counterexample. Then $(b;\rho;l;t)$ is an influence-tuple. Since \A\ is non-degenerate, there exists a non-degenerate interval $I$ such that for each $x\in I$
it holds that $(x,b_{-l};\rho;l;t)$ is an influence-tuple, and therefore a counterexample.
% Since \A\ is non-degenerate, there exists a non-degenerate interval $I$ such that for each $x\in I$ round $t$ is $(x,b_{-l};\rho)$-influential, and therefore  $(x, b_{-l}; \rho; l; t)$ is a counterexample.
Thus the set $S_l(b_{-l})$ in Proposition~\ref{prop:crux} cannot be finite, contradiction.

In the rest of this section we prove Proposition~\ref{prop:crux}. Fix a counterexample $(b; \rho; l; t)$; let $t'>t$ be the influenced round. In particular, $\A(b;\rho;t)=l$ (see \InFigureLAtRoundT{} in Figure~\ref{figure:proof-of-k-player-scale-free-iia} on page~\pageref{figure:proof-of-k-player-scale-free-iia}; all boxed numbers will refer to this figure). Then by the assumption there exist bids $b'$ such that $\A(b';\rho;t)=i'\neq l$.
We claim that this implies that there exists a bid $b^+_{i'}>b_{i'}$ such that $\A(b^+_{i'},b_{-i'};\rho;t)=i'$ (see $\InFigureIPrimeAtRoundTForLargeBid$).
This is proven in Lemma~\ref{lemma:a-player-can-increase-bid-and-get-impression} below, and in order to prove it we first present the following lemma, which essentially states that if the mechanism makes a choice between $i$ and $j$ of who to be show, then it can only depend on the ratio of their bids $\bid_i/\bid_j$, and not on the bids of other agents.

\begin{lemma}
Let $\A$ be an MAB (deterministic) allocation rule that is pointwise-monotone, scalefree, and satisfies IIA. Let there be two bid profiles $\alpha$ and $\beta$ such that $\A(\alpha;\rho;t)\in\{i,j\}$, $\A(\beta;\rho;t)\in\{i,j\}$, and $\alpha_i/\alpha_j = \beta_i/\beta_j$. Then it must be the case that $\A(\alpha;\rho;t) = \A(\beta;\rho;t)$.
\label{lemma:equal-ratio-of-bids-implies-same-allocation}
\end{lemma}
\begin{proof}
As \A\ is scalefree we assume that $\alpha_i=\beta_i$ and $\alpha_j=\beta_j$ by scaling bids in $\beta$ by a factor of $\alpha_i/\beta_i$ (or a factor of $\alpha_j/\beta_j$), without changing the allocation.

Assume for the sake of a contradiction that $\A(\beta;\rho;t) \neq \A(\alpha;\rho;t)$.
Let us number the agents as follows. Agents $i$ and $j$ are numbered $1$ and $2$, respectively. The rest of the agents are arbitrarily numbered $3$ to $k$. Consider the following sequence of bid vectors. $\alpha(1)=\alpha(2)=\alpha$ and $\alpha(m)=(\beta_m,\alpha(m-1)_{-m})$ for $m\in \{3,\ldots,k\}$. As $\alpha(1)=\alpha$ and $\alpha(k)=\beta$, $\A(\alpha(1);\rho;t) = \A(\alpha;\rho;t)$ and $\A(\alpha(k);\rho;t) = \A(\beta;\rho;t)$.
Since $\A(\alpha(k);\rho;t) = \A(\beta;\rho;t) \neq \A(\alpha;\rho;t)= \A(\alpha(1);\rho;t)$ there exists $m\in \{3,\ldots,k\}$ such that
$\A(\alpha(m-1);\rho;t) = \A(\alpha;\rho;t)\in \{i,j\}$ while $\A(\alpha(m);\rho;t)\neq \A(\alpha(m-1);\rho;t)$.
As $m\neq i$ and $m\neq j$, IIA implies that $\A(\alpha(m);\rho;t)=m$ and given that, IIA also implies that $\A(\alpha(k);\rho;t)\in \{m,m+1,\ldots k\}$ (note that $i,j$ are not in this set).
But as $\A(\alpha(k);\rho;t) = \A(\beta;\rho;t)\in \{i,j\}$  this yields a contradiction.
\end{proof}

% We present the following lemma which shows that the allocation of clicks between two agent only depends on the ratio of their bids, and is independent of the bids of the others.

\begin{lemma}
\label{lemma:a-player-can-increase-bid-and-get-impression}
Let $\A$ be an MAB (deterministic) allocation rule that is pointwise-monotone, scalefree, and satisfies IIA. Let there be two bid profiles $\alpha$ and $\beta$ such that $\A(\alpha;\rho;t)=i$ and $\A(\beta;\rho;t)=j\neq i$. Then there exists $\beta_i^+ > \beta_i$ such that $\A(\beta_i^{+},\beta_{-i};\rho;t)=i$.

In other words, if it is possible for $i$ to get the impression in round $t$ at all, then it is possible for her to get the impression starting from any bid profile and raising her bid high enough.
\end{lemma}
\begin{proof}
We first note that $\frac{\alpha_i}{\alpha_j}\geq \frac{\beta_i}{\beta_j}$. If not, then $\frac{\alpha_i}{\alpha_j} < \frac{\beta_i}{\beta_j}$. Consider a raised bid of $i$ from $\alpha_i$ to $\alpha^+_i=\alpha_j\cdot \frac{\beta_i}{\beta_j}$. In the bid profile $(\alpha^+_i,\alpha_{-i})$, $i$ must get the impression (by pointwise monotonicity). This gives a contradiction to Lemma~\ref{lemma:equal-ratio-of-bids-implies-same-allocation}, since $\A(\alpha^{+}_i,\alpha_{-i};\rho;t)=i\in\{i,j\}$, $\A(\beta;\rho;t)=j\in\{i,j\}$, and $\frac{\alpha_i^{+}}{\alpha_{j}}=\frac{\beta_i}{\beta_j}$, but $\A(\alpha^{+}_{i},\alpha_{-i};\rho;t)\not= \A(\beta;\rho;t)$.

Now, consider $i$ increasing her bid in profile $\beta$ to $\beta_{i}^{+}=\beta_j\cdot \frac{\alpha_i}{\alpha_j}$. Now, $\A(\alpha;\rho;t)=i\in\{i,j\}$, $\A(\beta_i^{+},\beta_{-i};\rho;t)\in\{i,j\}$ (from IIA), and $\frac{\alpha_i}{\alpha_j}=\frac{\beta_{i}^{+}}{\beta_j}$. We can apply Lemma~\ref{lemma:equal-ratio-of-bids-implies-same-allocation} to deduce that $\A(\alpha;\rho;t)=\A(\beta_{i}^{+},\beta_{-i};\rho;t)$ and both are equal to $i$ since the first allocation is equal to $i$.
\end{proof}

From the lemma above, it follows that agent $i'$ can increase her bid (in bid profile $b$) and get the impression in \realization $\rho$, round $t$. To quantify by how much agent $i'$ needs to raise her bid to get the impression, we introduce the notion of {\em threshold} $\Theta_{i,j}(\rho;t)$ in the next lemma.

\begin{lemma}
\label{lemma:well-defined-thetas}
Let \A\ be an MAB (deterministic) allocation rule that is pointwise monotone, scalefree and satisfies IIA.
For \realization $\rho$, round $t$, two agents $i$ and $j\not=i$, let bids $b_{-i-j}$ be such that there exist $x_0$ and $y$ satisfying $\A(x_0,y,b_{-i-j};\rho;t)=j$, and there exists $x$ (possibly dependent on $y$) satisfying $\A(x, y, b_{-i-j};\rho;t)=i$.
Let us fix such a $y$ and define\footnote{Note that if there are no values of bids of $i$ ($x_0$ and $x$) and $j$ (equal to $y$) such that $j$ can get an impression with small enough bid ($x_0$) of agent $i$ {\em and} $i$ can get an impression by raising her bid (to $x$), then we don't define $\Theta_{i,j}^{b_{-i-j}}(\rho;t)$ at all. We will be careful not to use such undefined $\Theta$'s. It is not hard to see that if bids are nonzero, then $\Theta_{i,j}(\rho;t)$ is defined if and only if $\Theta_{j,i}(\rho;t)$ is. Moreover $0<\Theta_{i,j}(\rho;t)<\infty$, and $\Theta_{j,i}(\rho;t)=(\Theta_{i,j}(\rho;t))^{-1}$.}
\begin{align*}
\Theta_{i,j}^{b_{-i-j}}(\rho,t) = \tfrac{1}{y}\inf_{x}\big\{x \bigm| \A(x,y,b_{-i};\rho;t)=i\big\}.
\end{align*}
Then for any bids $b'_{-i-j}$, $\Theta_{i,j}^{{b'_{-i-j}}}(\rho,t)$ is well defined and satisfies $\Theta_{i,j}^{{b'_{-i-j}}}(\rho,t) = \Theta_{i,j}^{b_{-i-j}}(\rho,t)$.
We denote it by $\Theta_{i,j}(\rho,t)$, as $\Theta_{i,j}^{b_{-i-j}}(\rho,t)$ is independent of $b_{-i-j}$.
%(Note that as \A\ is scalefree $\Theta_{i,j}(\rho,t)$ can only depend on the ratio of the bids of $i$ and $j$.)
\end{lemma}
\begin{proof}
We first prove that if the conditions of the definition of $\Theta_{i,j}^{b_{-i-j}}(\rho;t)$ are satisfied for $b_{-i-j}$, then are also satisfied for any other $b'_{-i-j}$. Let us say they are satisfied for $b_{-i-j}$, that is there exists $x_0$, $x$ and $y$, such that $\A(x_0,y,b_{-i-j};\rho;t)=j$ and $\A(x,y,b_{-i};\rho;t)=i$. We want to prove existence of $x'$ and $y'$ for $b'_{-i-j}$. If $\A(x_0,y,b'_{-i-j};\rho;t)=j$ then existence of $y'$ is proved for $b'_{-i-j}$ too, since $y'=y$ works. If not, then $\A(x_0,y,b'_{-i-j};\rho;t)=j'\not=j$ and $\A(x_0,y,b_{-i-j};\rho;t)=j$, and by Lemma~\ref{lemma:a-player-can-increase-bid-and-get-impression}, there exists a $y'>y$ such that $\A(x_0,y',b'_{-i-j};\rho;t)=j$. Once the existence of $y'$ is proved, we now prove the existence of $x'$. Let $x'=x\cdot \frac{y'}{y}\geq x$. We have $\A(x,y,b_{-i-j};\rho;t)=i\in\{i,j\}$ and $\A(x',y',b'_{-i-j};\rho;t)\in\{i,j\}$ by IIA ($i$ can only transfer impression to her by changing her bid) and $x'/y'=x/y$. From Lemma~\ref{lemma:equal-ratio-of-bids-implies-same-allocation}, we get $i=\A(x,y,b_{-i-j};\rho;t)=\A(x',y',b'_{-i-j};\rho;t)$. Hence the existence of $x'$ is proved too.

For the sake of contradiction, let us assume that $\theta:=\Theta_{i,j}^{b_{-i-j}}(\rho;t)<\Theta_{i,j}^{b'_{-i-j}}(\rho;t)=:\theta'$. Let us scale the bids in $(x',y',b'_{-i-j})$ by a factor such that the factor times $y'$ is equal to $y$. We can hence assume that $y'=y$. Let us pick a bid $x''\in (\theta y, \theta' y)$. We have $\A(x'',y,b_{-i-j};\rho;t)=i$ (since $x''/y$ is past the threshold $\theta$), $\A(x'',y'=y,b'_{-i-j};\rho;t)=j$ ($x''/y'$ is yet not past the threshold $\theta'$), and $x''/y = x''/y'$. This is a contradiction to the Lemma~\ref{lemma:equal-ratio-of-bids-implies-same-allocation}. Therefore, $\theta=\theta'$.
\OMIT{
To prove the claim it is sufficient to show that for any two bid vectors $b$ and $b'$ the following condition holds:
for any $\rho$ and $t$, if $\A(b;\rho;t)\in \{i,j\}$, $\A(b';\rho;t)\in \{i,j\}$ and $b_i/b_j = b'_i/b'_j$ then $\A(b';\rho;t) = \A(b;\rho;t)$. This can be seen by the following: if there are two bid vectors with $\Theta_{i,j}^{b}(\rho;t)\not=\Theta_{i,j}^{b'}(\rho;t)$, then we can find bid vectors $\bbar$ and $\bbar'$ such that $\A(\bbar;\rho;t)$ and $\A(\bbar';\rho;t)$ are both in $\{i,j\}$, $\bbar_i/\bbar_j=\bbar'_i/\bbar'_j$, and still $\A(\bbar;\rho;t)\not= \A(\bbar';\rho;t)$.
\yoginote{Moshe: do you agree with this statement that I added.}

As \A\ is scalefree we can scale all bids on $b'$ by factor $b_j/b'_j$ and get a new vector of bids $b''$ with $b''_j = b_j$
and $b''_i = b_i$ such that $\A(b'';\rho;t) = \A(b';\rho;t)$.

Assume in contradiction that $\A(b'';\rho;t) \neq \A(b;\rho;t)$.
Let us number the agents as follows. Agents $i$ and $j$ are numbered $1$ and $2$, respectively. The rest of the agents are arbitrarily numbered $3$ to $k$. Consider the following sequence of bid vectors. $b(1)=b(2)=b$ and $b(m)=(b''_m,b(m-1)_{-m})$ for $m\in \{3,\ldots,k\}$. As $b(1)=b$ and $b(k)=b''$, $\A(b(1);\rho;t) = \A(b;\rho;t)$ and $\A(b(k);\rho;t) = \A(b'';\rho;t)$.
Since $\A(b(k);\rho;t) = \A(b'';\rho;t) \neq \A(b;\rho;t)= \A(b(1);\rho;t)$ there exists $m\in \{3,\ldots,k\}$ such that
$\A(b(m-1);\rho;t) = \A(b;\rho;t)\in \{i,j\}$ while $\A(b(m);\rho;t)\neq \A(b(m-1);\rho;t)$.
As $m\neq i$ and $m\neq j$, IIA implies that $\A(b(m);\rho;t)=m$ and given that, IIA also implies that $\A(b(k);\rho;t)\in \{m,m+1,\ldots k\}$ (note that $i,j$ are not in this set).
But as $\A(b(k);\rho;t) = \A(b'';\rho;t)\in \{i,j\}$  this yields a contradiction.
}
\end{proof}

We conclude that if $b^+_{i'}> b_l\cdot \Theta_{i',l}(\rho,t)$ then $\A(b^+_{i'},b_{-i'};\rho;t)=i'\neq l$ (see $\InFigureIPrimeAtRoundTForLargeBid$ again). Note that we are using $\Theta_{i',l}(\rho;t)$ since this is well-defined.
Define $\rho' =\rho\xor\indicator(l,t)$.

Let us think about decreasing the bid of agent $l$ from $b_l$ (it is positive, since all bids are assumed to be positive). When the bid of agent $l$ is $b_l$, she gets the impression in round $t$, but when her bid is small enough (in particular as low as $b_{i'}/\Theta_{i',l}(\rho;t)$), then she must not get the impression in round $t$ (see Lemma~\ref{lemma:equal-ratio-of-bids-implies-same-allocation}). When the bid of $l$ decreases, some other agent gets the impression in round $t$, let us call that agent $i$ (note that this agent may not be the same as agent $i'$ above).
%\MBnote{Yogi, I do not understand this point: if the impression does not go to $i$ then $\Theta_{l,i}(\rho;t)$ is not well defined, but you use it later!}
%but agent $i'$ guarantees the existence of agent $i$).
See $\InFigureLDecreasesHerBid$.

Now, starting from bid profile $b$, let us increase the bid of agent $i$. When the bid of agent $i$ is large enough (in particular as large as $b_i \Theta_{i',l}(\rho;t) b_{l}/b_{i'}$), then $l$ can no longer get the impression in round $t$ (see Lemma~\ref{lemma:equal-ratio-of-bids-implies-same-allocation}). From IIA, the impression must get transferred to $i$. Therefore we can define $\Theta_{i,l}(\rho;t)$, and when $b_i^+ > b_{l}\Theta_{i,l}(\rho;t)$, agent $i$ gets the impression in round $t$ (see $\InFigureIIncreasesHerBid$ again). Note that $\A(b^+_i,b_{-i};\rho;t)=\A(b^+_i,b_{-i};\rho';t)=i$ (click information for $l$ at round $t$ cannot influence the impression decision at round $t$).

Recall that $t'$ is the influenced round.
%, and it is minimal.
Let $\A(b;\rho;t')=j$ and let $\A(b;\rho';t')=j'\neq j$ (see $\InFigureRoundTIsInfluential$).
As \A\ is pointwise monotone and IIA, $\A(b^+_i,b_{-i};\rho;t')\in \{i,j\}$ and $\A(b^+_i,b_{-i};\rho';t')\in \{i,j'\}$.
It must be the case that $\A(b^+_i,b_{-i};\rho;t')=\A(b^+_i,b_{-i};\rho';t')$, as $l$ does not get an impression at round $t$ (and the algorithm does not see the difference between $\rho$ and $\rho'$).
As $j'\neq j$ we conclude that
$$\A(b^+_i,b_{-i};\rho;t')=\A(b^+_i,b_{-i};\rho';t')=i.$$

Next we note that $i\neq j$ and $i\neq j'$. This is because if $i=j$ (respectively\ $i=j'$), then round $t$ would be $(b;\rho)$-influential (respectively $(b;\rho')$-influential) with influenced agent $i$ but it is not $(b;\rho)$-secured (respectively $(b;\rho')$-secured) from $i$, in contradiction to the assumption.

We also note that $l\in \{j,j'\}$ (see $\InFigureLIsJOrJPrime$). Assume for the sake of contradiction that $l\neq j$ and $l\neq j'$.
%(otherwise switch the names of $\rho$ and $\rho'$)
%we show that $l=j'$.
For $b^-_l < b_i\cdot \Theta_{l,i}(\rho,t)$ it holds that $\A(b^-_l,b_{-l};\rho;t)= \A(b^-_l,b_{-l};\rho';t)=i$ (since $i$ was defined such that $i$ gets the impression in round $t$ when $l$ decreases her bid) thus $\A(b^-_l,b_{-l};\rho;t') = \A(b^-_l,b_{-l};\rho';t')$ (as click information for $l$ at round $t$ is not observed). (Also, as a side note, observe that $b_l^{-}<b_l$ by pointwise-monotonicity since agent $l$ was getting an impression in round $t$ with bid $b_l$ and lost it when her bid is $b_l^{-}$.) Let $\A(b^-_l,b_{-l};\rho;t') = \A(b^-_l,b_{-l};\rho';t')=l'$. Note that $l'\not=l$, since otherwise, $\A_l(x,b_{-l};\rho;t')$ is not a monotone function of $x$: it is $0$ when $x=b_{l}$ (since $j$ gets an impression), and $1$ when $x=b_l^-<b_l$, a contradiction to pointwise-monotonicity.
Now, note that the impression in $\rho'$ at time $t'$ transfers from $j'$ to $l'$, and impression in $\rho$ at time $t'$ transfers from $j$ to $l'$, none of which ($\{j,j',l'\}$) are equal to $l$ and $j\not=j'$. Let us write this in equations:
\begin{align*}
\A(b_l,b_{-l};\rho;t')&=j  & \A(b_{l}^{-},b_{-l};\rho;t')&= l' \\
\A(b_l,b_{-l};\rho';t')&=j'  & \A(b_{l}^{-},b_{-l};\rho';t')&= l'.
\end{align*}
It must be the case that either $j\not=l'$ or $j'\not=l'$ (since $j\not= j'$). If $j\not=l'$, then in $\rho$ at time $t'$, reducing the bid of $l$ transfers impression from $j$ to $l'$ (both of them are different from $l$), thus violating IIA. Similarly, if $j'\not=l'$, then in $\rho'$ at time $t'$, reducing the bid of $l$ transfers impression from $j'$ to $l'$ (both of them are different from $l$), thus violating IIA. We thus have $l\in\{j,j'\}$. Let $l=j'$ (since otherwise, we can swap the roles of $\rho$ and $\rho'$).

\begin{comment}
This implies that by decreasing the bid of $l$,
at round $t'$ and impression moves from $j$ to $l$ in \realization $\rho$, and from $j'$ to $l$ in \realization $\rho'$.
As $l\neq j$ we get a contradiction to IIA unless $j=l$ and $j'=l$.
\end{comment}

To summarize what we have proved so far: there are 3 distinct agents $i,j,l$ such that
\begin{align*}
&\A(b;\rho;t)=\A(b;\rho';t)=\A(b;\rho';t')=l \quad\text{(since $\A(b;\rho';t')=j'=l$)},\\
&\A(b;\rho;t')=j \quad\text{and}\\
&\A(b^+_i,b_{-i};\rho;t)=\A(b^+_i,b_{-i};\rho;t')=\A(b^+_i,b_{-i};\rho';t)=\A(b^+_i,b_{-i};\rho';t')=i.
\end{align*}
Observe also that $\Theta_{i,l}(\rho,t)= \Theta_{i,l}(\rho',t)$ as $\rho$ and $\rho'$ only differ at a click at round $t$, and such a click cannot determine the allocation decision at round $t$.
Also, $\max \{\Theta_{i,j}(\rho,t')\cdot b_j, \Theta_{i,l}(\rho',t')\cdot b_l\} \le \Theta_{i,l}(\rho,t)\cdot b_l$ as the allocation at round $t'$, which is different for $\rho$ and $\rho'$ (at $b$), depends on $l$ getting the impression at round $t$.\footnote{In Figure~\ref{figure:proof-of-k-player-scale-free-iia} we defined $b_i^{+\rho}:= \Theta_{i,j}(\rho;t')b_j$ and $b_i^{+\rho'}:=\Theta_{i,l}(\rho';t')b_l$. These are the bids of agent $i$ at which impression transfers to her in round $t'$ in $\rho$ and $\rho'$ respectively. See $\InFigureBIPlusRho$ and $\InFigureBIPlusRhoPrime$ in the figure.} Finally we prove that
$\Theta_{i,j}(\rho,t')\cdot b_j$$= \Theta_{i,l}(\rho',t')\cdot b_l$ (see $\InFigureBIPlusRhoEqualsBIPlusRhoPrime$).

\begin{claim}
$\Theta_{i,j}(\rho,t')\cdot b_j= \Theta_{i,l}(\rho',t')\cdot b_l$
\end{claim}
\begin{proof}
First of all, note that $\Theta_{i,j}(\rho;t')$ and $\Theta_{i,l}(\rho',t')$ are well-defined. Let $\bar{b}_i= (\Theta_{i,j}(\rho,t')\cdot b_j+ \Theta_{i,l}(\rho',t')\cdot b_l) /2$.
Consider the following two cases.

If $\Theta_{i,j}(\rho,t')\cdot b_j <  \Theta_{i,l}(\rho',t')\cdot b_l$
then round $t$ is $(\bar{b}_i,b_{-i};\rho)$-influential (as $\A(\bar{b}_i,b_{-i};\rho;t') = i$ and $\A(\bar{b}_i,b_{-i};\rho';t') = l$) with influencing agent $l$ ($\A(\bar{b}_i,b_{-i};\rho;t)= \A(\bar{b}_i,b_{-i};\rho';t) = l$ since $\bar{b}_i < \Theta_{i,l}(\rho,t)\cdot b_l$) and influenced agent $i$.
Additionally, $t$ it is not $(\bar{b}_i,b_{-i};\rho)$-secured from $i$ (as $\A(b_i^{+},b_{-i};\rho;t) = \A(b_i^{+},b_{-i};\rho';t) = i$).
A contradiction to first condition in the theorem.

Similarly, if $\Theta_{i,j}(\rho,t')\cdot b_j >  \Theta_{i,l}(\rho',t')\cdot b_l$
then round $t$ is $(\bar{b}_i,b_{-i};\rho)$-influential (as now $\A(\bar{b}_i,b_{-i};\rho;t') = j$ and $\A(\bar{b}_i,b_{-i};\rho';t') = i$) with influencing agent $l$ and influenced agent $i$.
Additionally, $t$ it is not $(\bar{b}_i,b_{-i};\rho)$-secured from $i$.
Again, a contradiction to the first condition in the theorem.
\end{proof}

The lemma implies that $b_l\in S_l(b_{-l})$, where a finite set $S_l(b_{-l})$ is defined by
$$ S_l(b_{-l}) = \left\{
    b_j\; \frac{\Theta_{i,j}(\rho,t')}{\Theta_{i,l}(\rho',t')}:\;
        \text{all agents $i,j \neq l$, all \realizations $\rho, \rho'$ and all $t'$ s.t. $\frac{\Theta_{i,j}(\rho,t')}{\Theta_{i,l}(\rho',t')}$ is well-defined}
\right\}.
$$
This completes the proof of Proposition~\ref{prop:crux}.

%\input{app-characterization}
%\input{app-naive}
%\input{app-regret}
%\input{app-nonSF}
%\input{app-universal}
%\input{app-PSim}
%\input{app-expectation}

%AS2: removed it
%\input{app-ucb-counterexample}

\end{document}